\documentclass[pdflatex,sn-mathphys-num]{sn-jnl}

\usepackage{graphicx}%
\usepackage{multirow}%
\usepackage{amsmath,amssymb,amsfonts}%
\usepackage{amsthm}%
\usepackage{mathrsfs}%
\usepackage[title]{appendix}%
\usepackage{xcolor}%
\usepackage{textcomp}%
\usepackage{manyfoot}%
\usepackage{booktabs}%
\usepackage{algorithm}%
\usepackage{algorithmicx}%
\usepackage{algpseudocode}%
\usepackage{listings}%
\usepackage{multirow}

\newcommand{\chieff}{\chi_{\rm eff}}
\newcommand{\Msun}{{\rm M_\odot}}

\theoremstyle{thmstyleone}%
\theoremstyle{thmstyletwo}%

\theoremstyle{thmstylethree}%

\begin{document}

\title{Mass and spin properties of black-hole mergers reveal formation in triples}


\author*[1]{\fnm{Jakob} \sur{Stegmann}}\email{jstegmann@mpa-garching.mpg.de}

\author[2,1]{\fnm{Aleksandra} \sur{Olejak}}

\affil[1]{\orgname{Max Planck Institute for Astrophysics}, \orgaddress{\street{Karl-Schwarzschild-Str. 1}, \postcode{85748} \city{Garching}, \country{Germany}}}


\affil[2]{\orgname{Nicolaus Copernicus Astronomical Center, Polish Academy of Sciences}, \orgaddress{\street{Bartycka 18}, \postcode{00-716} \city{Warsaw}, \country{Poland}}}


\abstract{Deciphering the formation channels of the observed binary black hole mergers remains a central open problem in gravitational-wave astronomy. With hundreds of detections, the inferred distribution of masses and spins reveals a rich diversity which is difficult to reconcile with traditional formation scenarios from isolated binary stars or dense stellar environments. Here, we consider the large fraction of massive progenitor stars found in hierarchical triples and investigate black hole mergers driven by gravitational perturbations from tertiary companions. Simulations of their dynamics and stellar evolution reproduce key features of the merger population, including: i) a sharp primary mass peak at $m_1\sim10\,\rm M_\odot$ dominating the population; ii) substantial fractions of systems with large spin-orbit angles within and beyond the peak, matching the skewed distributions of the spin parameters $\chi_{\rm eff}$ and $\chi_p$; iii) a mass-ratio distribution favouring equal masses with matching slopes. We further predict a sharp decline of mergers near $m_1\sim30\,\rm M_\odot$, coincident with the location of inferred features that may signal another channel dominating the high-mass tail. Our models of hierarchical triples demonstrate, for the first time, a formation scenario which simultaneously reproduces the principal properties inferred for the low-mass bulk of binary black hole mergers.}

\maketitle

\section{Introduction}
Since the first direct detection of gravitational waves from a merging binary black hole \citep{GW150914}, the observational sample has grown to a few hundred events \citep{GWTC5}. Identifying the formation channels behind the observed merger events is essential for turning gravitational-wave observations into a robust probe of massive-star evolution, compact-object formation, and the environments in which binary black holes assemble across cosmic time \citep{Mapellibook2021}. However, distributions of black hole mass and spin properties inferred from the latest gravitational-wave catalogue \citep{GWTC5pop} challenge the two main formation scenarios proposed in the literature. On the one hand, interactions between isolated massive binary stars are expected to produce merging black holes with spins closely aligned with their orbit \citep[e.g.,][]{Belczynski2020} and favour lower masses due to binary stripping \citep{Belczynski2020, Olejak2021, vanSon2022,Maclean2026,Chen2026,BanerjeeOlejak2026,Broekgaarden2026,gallegosgarcia2026massratiodistributionlowmassbinary} and pair-instability supernovae \citep{Heger2003,Farmer2019,Farag2022}. On the other hand, binary black holes that dynamically assemble in dense stellar environments, e.g., the cores of globular clusters, are expected to coalesce with isotropically distributed spin-orbit angles \citep{Rodriguez2016,Farr2017,Farr2018,PhysRevLett.120.151101,Rodriguez2019} and generally favour heavier masses \citep{Rodriguez2016,Rodriguez2018cluster,DiCarlo2019, AntoniniGieles2023} that may populate the high-mass tail of the inferred population \citep[e.g.,][]{Antonini2026,Tong2026,Rinaldi2026,Plunkett2026}.

By contrast, the bulk of the inferred binary black hole population merges at relatively low masses encompassing a sharp, dominant peak around $\sim10\,\Msun$ and shows no evidence for strong spin-orbit alignment nor fully isotropic spin orientations. This is reflected in the inferred distribution of the effective spin parameter $\chi_{\rm eff}$ which is the best-constrained spin property in gravitational-wave observations:
\begin{equation}\label{eq:chieff}
\chieff=\frac{\chi_1\cos\theta_1+q\chi_2\cos\theta_2}{1+q},
\end{equation}
where $\chi_{1(2)}$ and $\theta_{1(2)}$ denote the spin magnitudes and spin-orbit tilt angles of the primary (secondary) black hole, respectively, and $q$ is the binary mass ratio. Both parametric and non-parametric population analyses consistently find that the $\chieff$ distribution around the $\sim10\,\Msun$ peak~---~as well as when marginalised over the whole population~---~exhibits a puzzling skewness with three characteristic features \citep{GWTC5pop,Ray2026,Cheng2026,Alvarez-Lopez2026,Rinaldi2026,Flanagan2026a,Flanagan2026b}: i) a peak at small positive values $\chieff\lesssim0.1$; ii) a majority of systems with $\chieff>0$; and iii) a substantial fraction of $>10\,\%$ with $\chieff<0$. Isolated binary formation channels, however, predict the $\chieff$ distribution to be very strongly dominated by positive values (due to a preference for aligned spins $\cos\theta_{1(2)}\approx1$) \citep{Farr2017,Gerosa2018,Belczynski2020, OlejakBelczynski2021,Kapil2026}, whereas dynamical formation scenarios predict it to be nearly symmetric about zero (due to a uniform distribution of $\cos\theta_{1(2)}$) \citep{Rodriguez2016,Farr2017,Farr2018,PhysRevLett.120.151101,Rodriguez2019}. Large, non-isotropic spin-orbit misalignment is also displayed in the distribution of individual tilt angles $\cos\theta_{1(2)}$ \citep{Stegmann2026,GWTC5pop}, albeit with much less significance due to greater measurement and population inference uncertainties \citep{VitaleMould2025,Wolfe2026,wolfe2026binaryblackholespinpopulation}.

\begin{figure}
    \centering
    \includegraphics[width=\linewidth]{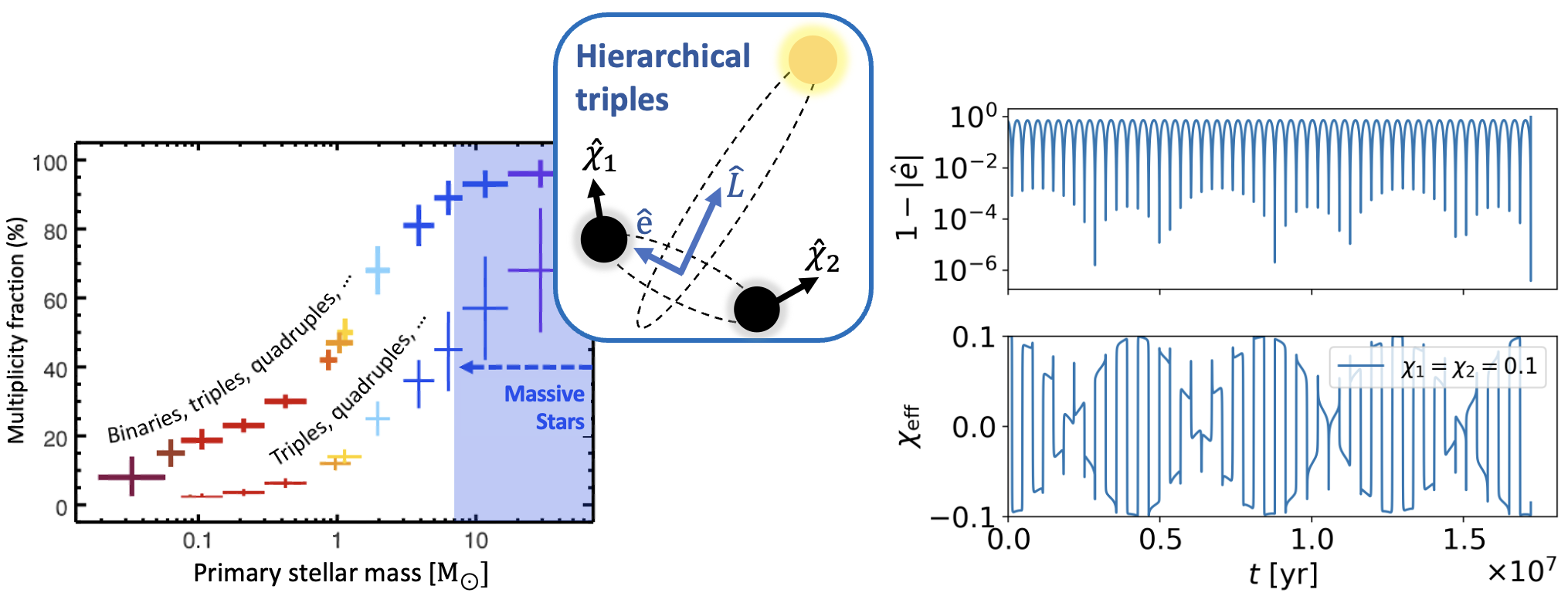}
    \caption{Left: Observed stellar multiplicity as a function of the mass of the most massive star (the primary) in a system (adapted from Ref.~\cite{Offner2023}). Most massive black hole progenitor stars are found in hierarchical triples or higher-order configurations, where a close inner binary is orbited by one or more distant companions. Right: A tertiary companion drives the merger of a binary black hole formed in the inner orbit (inset sketch), by inducing large-amplitude ``Lidov-Kozai" oscillations of the inner binary eccentricity vector $\hat{e}$. This process can be accompanied by large oscillations of the tilt angles $\cos\theta_{1(2)}=\hat{\chi}_{1(2)}\cdot\hat{L}$ between the black hole spin unit vectors $\hat{\chi}_{1(2)}$ and the inner binary orbital angular momentum unit vector $\hat{L}$, yielding characteristic values for $\chieff$; cf. Eq.~\eqref{eq:chieff}. The triple parameters of the simulated merger are chosen to reproduce the example shown in Fig.~2 in Ref.~\citep{Rodriguez2018}, but with black hole spin magnitudes set to $\chi_1=\chi_2=0.1$.}
    \label{fig:sketch}
\end{figure}

Electromagnetic surveys of massive stars may offer an important clue to resolving this problem. Most black-hole progenitor stars are found to reside in hierarchical triples or higher-order configurations, where a close inner binary is orbited by one or more distant companions \citep{Moe2017,Offner2023,Bordier2026}. In hierarchical triples, the gravitational perturbation from tertiary companions can induce large-amplitude ``Lidov–Kozai" oscillations \citep{Lidov1962,Kozai1962} of the eccentricity and orbital orientation of black holes formed in the inner binary. This process can cause mergers during close pericentre passages \citep{Silsbee2017,Antonini2017,Grishin2018,Liu2018,Antonini2018,Rodriguez2018,Mangipudi2022,Stegmann2022,2025A&A...699A.272V,Dorozsmai2025,Stegmann2025,Bruenech2026} and induce a wide range of characteristic spin-orbit angles \citep{Antonini2018,Liu2018,Su2021} (Fig.~\ref{fig:sketch}). Yet, despite the ubiquity of tertiary and higher-order companions in black-hole progenitor systems, their impact is neglected in most formation models of gravitational-wave sources and population inference in the data.

\section{Results}\label{sec2}

\begin{figure*}
\centering
\includegraphics[width=\linewidth]{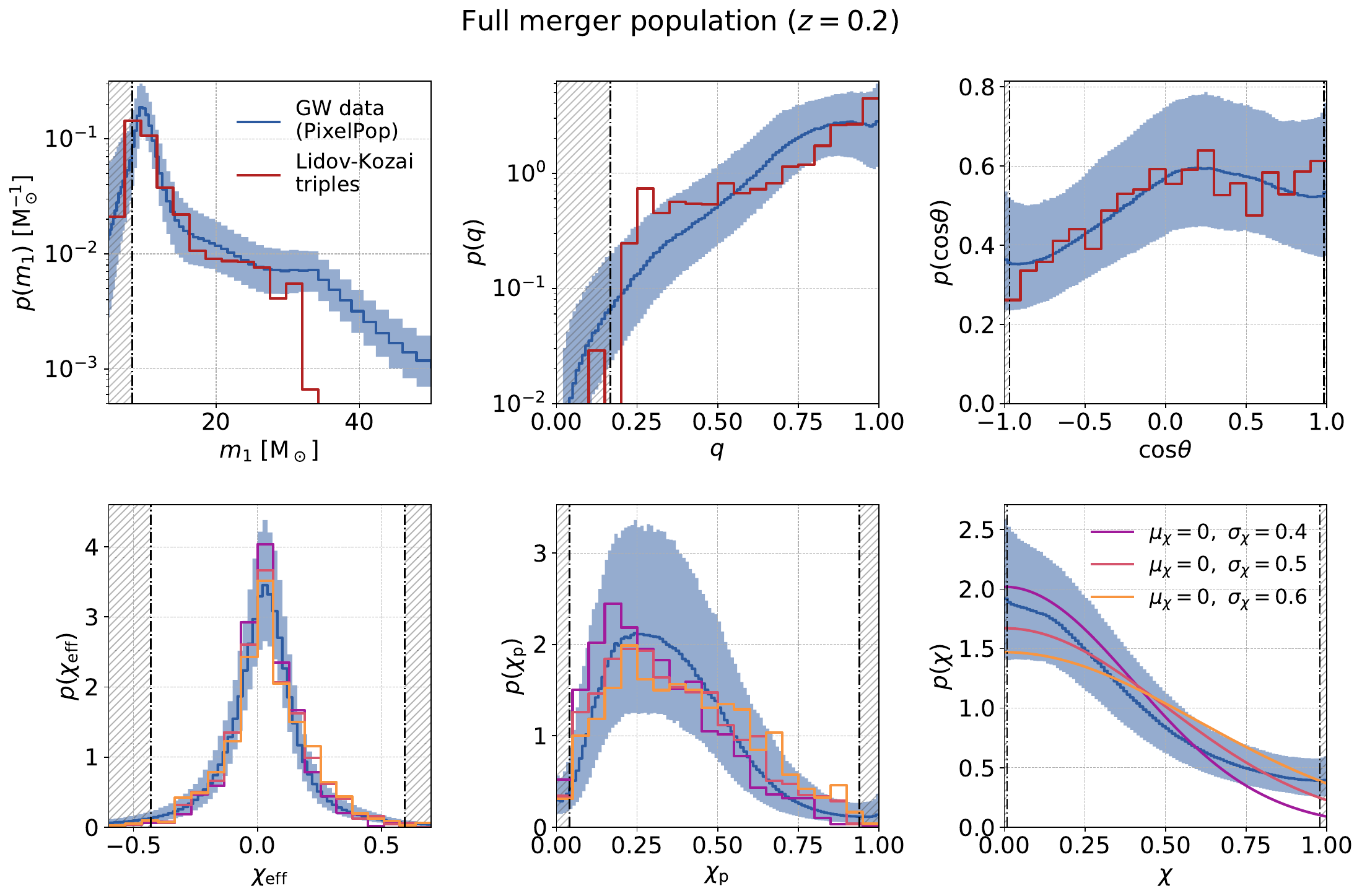}
\caption{Comparison between the properties of binary black hole mergers formed in hierarchical triples and the marginalised distributions inferred from gravitational-wave data. All properties are shown at a merger redshift $z=0.2$. The solid blue line shows the median merger-rate density inferred for GWTC-5.0 with the \texttt{PixelPop} model \cite{GWTC5pop}, with the shaded bands depicting $90\%$ credible regions. Hatched regions indicate parts of the parameter space that may not be strongly constrained by the data \cite{GWTC5pop}. In the bottom row, we assume that both black-hole spin magnitudes are independently drawn from truncated normal distributions between $\chi=0$ and 1 with $\mu_\chi$ and $\sigma_\chi$ specified in the legend.}\label{fig:All-mergers}
\end{figure*}


Rather than asking whether hierarchical triples can produce merging black holes, we ask whether they explain the population properties inferred from the latest gravitational-wave observations \citep{GWTC5pop}. To test this, we simulate $3.2\times10^8$ hierarchical massive stellar triples from the zero-age main sequence (ZAMS) through stellar evolution and compact-object formation, and follow the surviving black-hole triples to gravitational-wave merger. We model the stellar evolution of the interacting inner binary and tertiary companion with the \texttt{COMPAS} stellar-evolution code \citep{COMPAS:2021,COMPAS:2025}, and their subsequent post-Newtonian gravitational dynamics with the secular three-body code presented in Refs.~\cite{Antonini2017,Rodriguez2018,rodriguez2019kozai}. Details of the numerical setup and adopted stellar-evolution prescriptions are provided in the Methods (Sec.~\ref{sec4}).

Fig.~\ref{fig:All-mergers} compares the resulting parameter distributions of binary black hole mergers from our triple simulations with the astrophysical merger population inferred from the latest gravitational-wave catalogue GWTC-5.0 \cite{GWTC5}. For the latter, we show the public result from the \texttt{PixelPop} model \cite{GWTC5-zenodo}, which is the default weakly-parametrised model of the LIGO-Virgo-KAGRA population analysis with minimal a priori assumptions about the underlying astrophysical population \cite{GWTC5pop}. Fig.~\ref{fig:All-mergers} shows that our default triple model produces mergers that broadly recover the inferred distributions of mass and spin properties. In particular, we find:

\begin{itemize}
    \item The primary mass distribution (upper left panel) reproduces the sharp peak at $m_1\approx10\,\Msun$,  with slopes on both sides consistent with the inferred distribution.
    \item The primary mass distribution exhibits a sharp drop near $m_1\approx30\,\Msun$.
    \item The mass ratio distribution $q=m_2/m_1$ of the secondary black-hole mass $m_2$ to the primary black-hole mass $m_1$ (top middle panel) favours equal-mass mergers ($q\approx1$) and broadly reproduces the inferred decline down to $q\approx0.2$.
    \item The distribution of individual black hole spin-orbit angles $\cos\theta$ (upper right panel) shows a preference for some degree of alignment $\cos\theta\gtrsim0$ and features mild evidence for peaks between $\cos\theta\approx0$ and $0.5$. We find that $\approx57\,\%$ of black holes have spin-orbit angles $\cos\theta>0$, in agreement with $57\pm5\,\%$ inferred from the data \citep{GWTC5pop}. The residual fraction displays very large spin-orbit angles ($\cos\theta<0$).
\end{itemize}

Previously, we have shown that the spin-orbit tilt distribution from triple simulations \cite{Antonini2017} is consistent with the observationally inferred distribution from the earlier catalogue GWTC-4 \cite{GWTC4pop}, but were unable to report large significance due to the great uncertainty of individual spin-orbit tilt measurements \cite{Stegmann2026}. Here, we find that the characteristic spin-orbit tilt distribution from triples shapes the better measured effective spin parameters $\chi_{\rm eff}$ and $\chi_p$ in a way that reproduces the observed distributions. To this end, we sample the individual black hole spin magnitudes independently from a truncated normal distribution between $\chi=0$ and 1 and explore three different variants for the distribution parameters $(\mu_\chi,\sigma_\chi)=(0.0,0.4)$, $(0.0,0.5)$, and $(0.0,0.6)$. The functional form of the spin magnitude distribution is heuristically chosen to reproduce the inferred distribution from the observations (bottom right panel). In particular, these spin magnitude models are consistent with a preference for small spin magnitudes and an extended tail to larger values. Ref.~\cite{GWTC5pop} reports that $73\pm0.4\,\%$ of black holes merge with spin magnitudes $\chi\leq0.5$; our chosen spin models with $\mu_\chi=0$ are designed to broadly recover this range with cumulative fractions $P(\chi\leq0.5; \sigma_\chi=0.4)\approx79.9\,\%$, $P(\chi\leq0.5; \sigma_\chi=0.5)\approx71.5\,\%$, and $P(\chi\leq0.5; \sigma_\chi=0.6)\approx65.8\,\%$. Using these spin magnitude models, we find:

\begin{itemize}
    \item The effective spin parameter distribution (bottom left panel) peaks in all three magnitude models at a small positive value $\chi_{\rm eff}<0.1$, in agreement with the observational inference. The triple simulations recover the inferred skewed shape of $\chieff$, which is simultaneously characterised by a majority with $\chi_{\rm eff}>0$ but a substantial fraction with $\chi_{\rm eff}<0$. Ref.~\cite{GWTC5pop} infers the degree of asymmetry about zero through the difference of cumulative fractions $\delta\chi=P(\chi_{\rm eff}>0)-P(\chi_{\rm eff}<0)$ and finds $\delta\chi=0.28^{+0.11}_{-0.12}$. Here, we find a matching $\delta\chi\approx0.24$ to 0.25 for all adopted $\sigma_\chi$. Equivalently, Ref.~\cite{GWTC5pop} infers that $36\pm6\,\%$ of mergers have negative $\chi_{\rm eff}$, consistent with our inferred fraction of $\approx38\,\%$ for all $\sigma_\chi$. 
    \item Agreement persists with the effective spin precession parameter $\chi_p=\max\left[\chi_{1\perp},\,\chi_{2\perp}q(4q+3)/(4+3q)\right]$ that quantifies the spin components $\chi_{1(2)\perp}$ along the orbital plane \citep{Schmidt2015}, although it is less precisely measured than $\chi_{\rm eff}$ (bottom middle panel). The inferred distribution and our models peak at around $\chi_p\approx0.2$ to 0.3 and show an extended tail to high values, further indicating evidence for large spin-orbit angles in the population.
\end{itemize}

\begin{figure*}
\centering
\includegraphics[width=\linewidth]{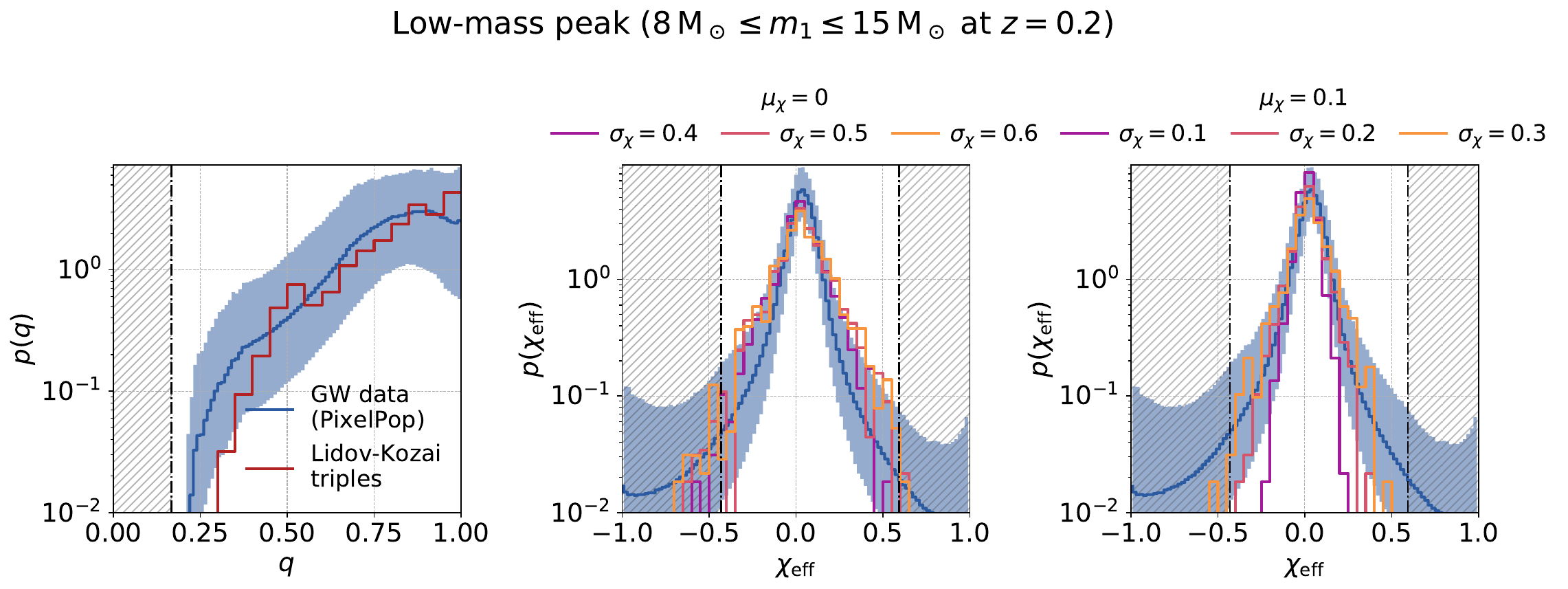}
\caption{Mass ratio (left panel) and effective spin parameter distributions (middle and right) of mergers with $8\,\Msun\leq m_1\leq15\,\Msun$. As in Fig.~\ref{fig:All-mergers}, the effective spin parameter distribution rests on the assumption of truncated normal distributions for the individual spin magnitudes with parameters $\mu_\chi$ and $\sigma_\chi$  shown in the legend.}\label{fig:Peak-mergers}
\end{figure*}

As shown in the upper left panel, triples may fully dominate the formation of mergers around the global peak at $m_1\sim10\,\Msun$. To further test this hypothesis we focus in Fig.~\ref{fig:Peak-mergers} on the subset of mergers with primary masses $8\,\Msun\leq m_1\leq15\,\Msun$. The left panel shows that in this mass range, the predicted mass ratio distribution closely follows the median inferred from the observations. To compare the $\chi_{\rm eff}$ distributions we explore additional spin magnitude models with $(\mu_\chi,\sigma_\chi)=(0.1,0.1)$, $(0.1,0.2)$, and $(0.1,0.3)$ (right panel) to account for the possibility of a global peak at $\chi\approx0.1$ inferred for the $\sim10\,\Msun$ peak (see lower left panel of Fig.~6 in Ref.~\cite{GWTC5pop}). All adopted spin magnitude models agree well with the observed distribution of $\chi_{\rm eff}$ and mostly differ in the tails of the distribution. We find that $(\mu_\chi,\sigma_\chi)=(0.1,0.2)$ fits the data particularly well with a median $\chi_{\rm eff}\approx0.012$ in agreement with $\chi_{\rm eff}=0.05^{+0.15}_{-0.19}$ inferred within the mass range \cite{GWTC5pop}. The median values in our other spin models range from 0.01 to 0.03. 

Our models also support a reported tendency towards a broader and more symmetric $\chi_{\rm eff}$ distribution beyond the peak ($m_1>15\,\Msun$) than within ($8\,\Msun\leq m_1\leq15\,\Msun$) \citep{GWTC5pop,Flanagan2026a,Rinaldi2026}. In all spin models, we find the cumulative fraction $P(\chi_{\rm eff}<0)$ to grow to $\approx42\%$ at $m_1>15\,\Msun$ and the median $\chi_{\rm eff}$ value to shrink to $\approx0.01$ to $0.026$, consistent with an inferred median $\chi_{\rm eff}=0.01^{+0.43}_{-0.52}$ \cite{GWTC5pop}. However, we stress that this comparison is necessarily approximate if other formation channels dominate the high-mass tail of the distribution $(m_1\gtrsim30\,\Msun)$ and further affect the shape of the $\chi_{\rm eff}$ distribution \citep[e.g.,][]{Antonini2026,Rinaldi2026}.

As discussed in the Methods (Sec.~\ref{sec4}), we discard mergers with short delay times $t_d<t_{\rm th}$ between ZAMS and the formation of gravitational-wave mergers. In that way, we conservatively exclude black-hole mergers which, in reality, may instead have been driven to a stellar merger before compact-object formation. As a default, we assume a threshold $t_{\rm th}=5\,\rm Myr$ which is a typical stellar lifetime of massive stars. Using this threshold and the cosmological model from Ref.~\cite{Neijssel2019} described in the Methods, we infer binary black hole merger rates ranging from $2.6$ to $5.3\,\rm Gpc^{-3}\,yr^{-1}$ at a reference redshift $z=0.2$ across the model variations described in the Supplementary Material~\ref{sec:models}. The predicted rate roughly doubles if we include mergers with shorter delay times (i.e., $t_{\rm th}=0\,\rm Myr$), reaching $5.8$~---~$11.5\,\rm Gpc^{-3}\,yr^{-1}$. These rates constitute significant fractions of the range $27.5$~---~$49.4\,\rm Gpc^{-3}\,yr^{-1}$ ($90\,\%$ credible interval) inferred for the total merger population from gravitational-wave observations \citep{GWTC5pop}. We emphasise, however, that theoretical merger-rate predictions are subject to substantial uncertainties associated with the underlying assumptions about stellar evolution, compact-object formation, cosmic star-formation and metallicity histories, and, in our case, the initial properties of triple systems. Indeed, predicted binary black hole merger rates vary considerably across models in the literature \citep[cf.,][]{MandelBroekgaarden2022}, with a tendency to overpredict the observationally inferred rate, in some cases by a few orders of magnitude \citep{Boco2026,Broekgaarden2026b}. In this context, agreement within a factor of a few is encouraging, although future work is needed to assess the variance and robustness of triple-driven merger rates across different cosmological and stellar-evolution models. Restricting the mergers to $8\,\Msun\leq m_1\leq15\,\Msun$, we find $1.7$~---~$3.2\,\rm Gpc^{-3}\,yr^{-1}$ for $t_{\rm th}=5\,\rm Myr$ and $3.9$~---~$7.9\,\rm Gpc^{-3}\,yr^{-1}$ without any threshold.

While we find the parameter distributions (Figs.~\ref{fig:All-mergers} and~\ref{fig:Peak-mergers}) to be broadly consistent with and without restricting the population to $t_{\rm th}=5\,\rm Myr$, the number of eccentric mergers substantially differs and may therefore be used as a discriminator to study the impact of three-body dynamics during the stellar evolution of triples. In Fig.~\ref{fig:eccentricity} (left panel) of the Supplementary Material~\ref{sec:models} we show the eccentricity distribution of mergers at $z=0.2$ and at a dominant gravitational-wave frequency of $10\,\rm Hz$. We find that the fraction of mergers with potentially measurable eccentricity ($e_{\rm 10\,Hz}>0.1$) is zero for  $t_{\rm th}=5\,\rm Myr$ and $3.0\,\%$ for $t_{\rm th}=0\,\rm Myr$. Thus, the detection of eccentric sources in the mass range of triple-driven mergers may indicate a sufficient number of triples with short delay times that survived stellar evolution. Figure~\ref{fig:eccentricity} (right panel) shows that, although a significant fraction of eccentric mergers exhibit large spin--orbit misalignments, the distribution is dominated by closely aligned spins ($\cos\theta\approx1$). These binaries are rapidly driven to highly eccentric mergers before their orbital angular momentum can tilt substantially away from the spin directions \citep{Stegmann2025}. This contrasts with eccentric binaries assembled dynamically in dense environments, whose spin orientations are expected to be isotropic \citep{StegmannGerosa2025}. The distinction may become observable as waveform models and searches that jointly incorporate eccentricity and spin precession mature \citep[e.g.,][]{Liu:2024SEOBNRE,Morras:2025nlp,Albanesi:2025txj,Morras:2026pyEFPEHM,Gamboa:2026SEOBNRv6EPHM}. Detections of low-mass, spin-aligned eccentric events could therefore provide a distinctive signature of short-delay triple mergers.

\section{Discussion}\label{sec3}

In our models, the sharpness of the $\sim 10\,\Msun$ peak results from the combination of the adopted rapid core-collapse engine \citep{Fryer2022} and short delay times characteristic of triple-driven mergers. The latter lead to a dominant contribution of high-metallicity progenitors to the local merger rate, limiting black-hole masses through efficient binary stripping and strong stellar winds. The choice of core-collapse engine also plays an important role, as it determines the characteristic threshold for direct black-hole formation and the relative abundance of lower-mass black holes. The rise toward the peak at $\sim 10\,\Msun$ has also been proposed in the literature to be linked to the compactness structure of pre-supernova stellar cores \citep{Schneider2023}, but the steep decline above the $\sim 10\,\Msun$ peak has not been reproduced by other proposed formation channels in the literature. In isolated binary evolution, highly non-conservative mass transfer has been identified as a possible mechanism for shifting the mass distribution toward the observed $\sim 10\,\Msun$ peak, whereas more conservative mass-transfer prescriptions tend to produce a larger contribution from higher-mass primary black holes \citep{Maclean2026,Chen2026,Olejak2026b}. Nevertheless, these models tend to over-predict the merger rate above the $\sim 10\,\Msun$ peak, even across a wide range of assumptions about the cosmological star formation history \citep{vanSon2023}.

Although we adopt highly non-conservative mass transfer as our default assumption, we show in the Supplementary Material~\ref{sec:models} that reproducing the sharp $\sim 10\,\Msun$ peak in the triple channel does not require a particular prescription for the uncertain mass-transfer physics. The peak persists across alternative models considered here because its origin is primarily associated with the enhanced contribution of high-metallicity progenitors to the local merger population, resulting from the short delay times of triple-driven mergers. Our default treatment of binary stripping nevertheless differs from that adopted in many previous studies in its prescription for angular-momentum loss. This choice is motivated by recent observations of Gaia black holes in binaries with stellar companions \citep{ElBadry2023a,ElBadry2023b,Chakrabarti2023}. Their properties are difficult to reproduce under standard binary-interaction models assuming isotropic re-emission and instead favour scenarios in which the non-accreted material carries approximately the specific orbital angular momentum of the donor \citep{Olejak2026a,Mapelli2026,Xu2026}.

Alternative ways to reproduce the inferred trends of the spin observables may require non-standard assumptions about compact-object formation or some degree of fine-tuning. First, the $>10\,\%$ fraction of systems with negative $\chieff$ in the $\sim10\,\Msun$ peak may be understood as a mixture of an isolated binary formation scenario (reproducing the overall preference for positive $\chieff$) and dynamical binary formation in star clusters (reproducing the fraction of systems with negative $\chieff$). This would require both channels to contribute to the $\sim10\,\Msun$ peak at closely matching merger rates, which may be interpreted as a classical fine-tuning problem \citep{Barnes2018} given the wide range of total merger rates possible in each scenario \citep{MandelBroekgaarden2022}. In particular, simulations of low-mass mergers in young star clusters have been shown to depend sensitively on cluster formation histories in metal-rich environments at low redshift \citep{DiCarlo2020,Ye2026}. Meanwhile, scenarios that have been proposed for the $\sim10\,\Msun$ peak \citep[e.g.,][]{vanSon2023,Chen2026,Ye2026} do not reproduce its inferred sharpness \citep{Ray2026b}. Second, a significant fraction of systems with negative $\chieff$ may be possible in the isolated binary formation scenario if black holes receive very large natal kicks ($>100\,\rm km/s$) at formation that would flip the binary orbit with respect to the spins \citep{Gerosa2018, Olejak2024}. As discussed in Ref.~\citep{Flanagan2026b}, it is highly uncertain whether black holes around the $\sim10\,\Msun$ peak routinely receive such large natal kicks. Simultaneously producing a large kick and a $\sim10\,\Msun$ black hole requires an energetic, asymmetric explosion that ejects sufficient momentum while retaining enough fallback. Current three-dimensional supernova simulations yield both low- and high-kick outcomes, sometimes for the same progenitor \citep{2024Ap&SS.369...80J,2025ApJ...987..164B}, leaving their occurrence rate uncertain. Observational constraints are similarly mixed: some black holes appear to receive small kicks \citep[e.g.,][]{2024PhRvL.132s1403V,whitaker2026longperiodstellarmassblack}, whereas evidence for kicks $\gtrsim100\,\rm km/s$ in low-mass X-ray binaries remains disputed \citep[][and references therein]{Mandel2016}. Third, a significant fraction of systems with negative $\chieff$ may be obtained if the spatial direction of black hole natal kicks is not random but exhibits some correlation with the spin direction of the remnant \citep{BaibhavKalogera2024}, or if black holes do not inherit the spin direction of their progenitor stars \citep{Tauris2022}.

In this work, we have studied a formation scenario to reproduce gravitational-wave observations from the evolution of hierarchical triples without relying on any of these assumptions. Instead, we have shown that the most common gravitational-wave mergers can arise from the most commonly observed configuration of massive black-hole progenitor stars. 

\section{Methods}\label{sec4}
We simulate populations of hierarchical stellar triples from their ZAMS through compact-object formation and gravitational-wave merger. Each population consists of 32 metallicity bins that are log-uniformly distributed between $Z=0.0001$ and $0.03$. In each bin we simulate $10^7$ triples, summing up to $3.2\times10^8$ in total.  The stellar triples are initiated as follows. For the inner binary, we sample the mass $m_1$ of the primary star from the Kroupa initial mass function $p(m_1)\propto m_1^{-2.3}$ between $5$ and $150\,\Msun$ \citep{Kroupa}. Adopting the observationally inferred distribution of  massive binary stars by \citet{Sana2012}, we sample the inner binary mass ratio $q_1=m_2/m_1$ from a power-law $p(q_1)\propto q_1^{-0.1}$ between 0.1 and 1.0 and its orbital period from $p(\log P_1/{\rm days})\propto(\log P_1/{\rm days})^{-0.55}$, limiting the resulting semi-major axis to $10^{-2}$ to $10^3\,\rm AU$. For the eccentricity distribution, we assume $p(e_1)\propto e_1^{-0.45}$ between zero and one \citep{Sana2012} and reject any inner binary which would overflow its Roche-lobe at ZAMS. For the outer orbits of the tertiary companions around the inner binary barycentres, we sample the outer semi-major axis $a_2$ from a log-uniform distribution in $a_2/a_1$ between 1 and $10^3$, the outer eccentricity $e_2$ from a thermal distribution $p(e_2)=2e_2$, and the outer mass ratio $q_2=m_3/(m_1+m_2)$ from a power-law $p(q_2)\propto q_2^{0.94}$ between 0.1 and 1.0, as inferred from recent parameter distributions of main-sequence massive triple observations in the SMaSH+ survey \citep{Bordier2026}. We further require the tertiary orbit to satisfy the dynamical stability criterion given in Ref.~\cite{Mardling}
\begin{equation}
    a_2>2.8a_1\left[\left(1+\frac{m_3}{m_1+m_2}\right)\frac{1+e_2}{(1-e_2)^3}\right]^{2/5}\label{eq:stability},
\end{equation}
and impose an upper limit of $a_2\leq10^{4}\,\rm AU$, beyond which binaries likely disrupt due to stellar fly-bys and tides from the host galaxy \citep{Jiang,El-Badry2018,Hamilton2024,Stegmann2024}.

The stellar evolution of triples is modelled with the rapid population synthesis code \texttt{COMPAS} (version v03.29.05) \citep[][]{COMPAS:2021,COMPAS:2025}, using its binary mode (``BSE") for the inner binary stars and its single star mode (``SSE") for the tertiary companion. While \texttt{COMPAS} self-consistently treats the evolution of interacting binary stars, including mass-transfer episodes, wind mass loss, and compact-object formation, we account for the evolution of the outer orbit through orbital widening caused by mass loss and orbital changes caused by compact-object formation in the inner binary or tertiary companion, as in Refs.~\citep{Stegmann2022,LuNaoz}. Throughout the entire evolution, we continuously check for instability according to Eq.~\eqref{eq:stability}, stellar mergers in the inner binary, and orbital disruptions at black-hole formation and terminate the simulation if any of these occurs. In our default model, we assume fully non-conservative mass transfer, with all transferred mass lost from the system carrying the specific orbital angular momentum of the donor. However, as demonstrated in the Supplementary Material~\ref{sec:models}, variations in the adopted mass-transfer physics do not significantly affect the merger rates, masses, or spins of the resulting binary black hole mergers. Core-collapse supernova remnants are determined following prescriptions from Ref.~\citep{Fryer2022}, adopting a mixing parameter of \(f_{\rm mix}=4.0\) (rapid-type explosion model), and a critical pre-collapse core-mass threshold of \(M_{\rm crit} = 4.75\,\Msun\) for black-hole formation. 

Triples that survive the stellar evolution and form an inner binary black hole are integrated with the secular three-body code presented in Refs.~\citep[][]{Antonini2018,Rodriguez2018,rodriguez2019kozai}. This method employs the double-averaged equations of motion of hierarchical Newtonian three-body systems up to the octupole order \citep[e.g.,][]{Naoz2013} and includes post-Newtonian (pN) corrections for the inner orbit up to 2.5pN order and black-hole spin-orbit and spin-spin coupling to 2.0pN order. Simulations are performed up to a maximum integration time $t_{\rm max}={\rm min}[10^3\times t_{\rm LK},13.8\,\rm Gyr]$, where $t_{\rm LK}$ refers to the Lidov-Kozai timescale of the black hole triples \citep{Rodriguez2018}. Binary black hole mergers are recorded if the dominant gravitational-wave frequency \citep{Wen2003},
\begin{equation}
    f_{\rm GW}(e_1)=\frac{2f_{\rm orb}(1+e_1)^{1.1954}}{(1-e_1^2)^{1.5}},\label{eq:fGW}
\end{equation}  
enters the detector bandwidth at $10\,\rm Hz$, where $f_{\rm orb}$ is the orbital frequency of the binary. Our reported quantities $\chieff$, $\chi_p$, and $\cos\theta_{1(2)}$ are evaluated at this reference frequency. 

Thus, for each metallicity bin we obtain a finite number of gravitational-wave mergers characterised by mass and spin properties. To convert this synthetic merger sample into an astrophysical merger population at redshift $z$ we weight the mergers by accounting for the progenitor metallicity and delay times $t_d$ between ZAMS and gravitational-wave merger. To this end, we follow Ref.~\citep{Neijssel2019} (and its computational implementation in \texttt{COMPAS}) and assume that the cosmic star-formation-rate density is parametrised as
\begin{equation}
\psi(z)
=
a_{\rm SF}
\frac{(1+z)^{b_{\rm SF}}}
{1+\left[(1+z)/c_{\rm SF}\right]^{d_{\rm SF}}},
\end{equation}
where $a_{\rm SF}=0.01\,{\rm \Msun\,yr^{-1}\,Mpc^{-3}}$, $b_{\rm SF}=2.77$, $c_{\rm SF}=2.9$, and $d_{\rm SF}=4.7$. For the chemical evolution, we assume that the metallicity-specific star-formation rate separates into
\begin{equation}
\frac{{\rm d}^{3}M_{\rm SF}}{{\rm d}t_{\rm s}\,{\rm d}V_{\rm c}\,{\rm d}Z}
=
\psi(z)\,p(Z\mid z),
\end{equation}
where the metallicity distribution at fixed redshift is lognormal,
\begin{equation}
p(Z\mid z)
=
\frac{1}{Z\sigma\sqrt{2\pi}}
\exp\left[
-\frac{\left(\ln Z-\mu(z)\right)^2}{2\sigma^2}
\right].
\end{equation}
Its mean evolves with redshift according to
\begin{equation}
\langle Z(z)\rangle
=
\exp\left[\mu(z)+\frac{\sigma^2}{2}\right]
=
Z_0\,10^{\alpha_Z z},
\end{equation}
or equivalently
\begin{equation}
\mu(z)
=
\ln\left(Z_0\,10^{\alpha_Z z}\right)
-\frac{\sigma^2}{2}.
\end{equation}
We adopt $Z_0=0.035$, $\alpha_Z=-0.23$, and a redshift-independent width $\sigma=0.39$ in $\ln Z$, with no skewness. In the notation of the \texttt{COMPAS} cosmic integrator, these values correspond to $(\mu_0,\mu_z,\sigma_0,\sigma_z,\alpha_{\rm skew})=(0.035,-0.23,0.39,0,0)$.

There are several approximations involved in our integrations to reduce computational costs and enable simulations of large triple populations: i) The secular method formally breaks down if the timescale of orbital changes becomes shorter than the periods of the inner or outer orbit. Direct three-body integrations of hierarchical triples exhibit a tendency to produce mergers more efficiently than secular methods, particularly for mergers that enter the detector bandwidth with residual eccentricity \citep{Antonini2014}. Therefore, our findings likely underestimate the number of triple-driven mergers; ii) We account for the gravitational three-body dynamics only after the formation of black holes. Beforehand, the Lidov-Kozai effect is not taken into account, although we do consider orbital changes due to mass-loss and compact-object formation as explained above. This neglects the possibility of tertiary-driven Roche-lobe overflow or stellar mergers in the inner binary \citep{Stegmann2022,Stegmann2022b}. Taking these effects into account requires a simultaneous treatment of the Lidov-Kozai effect and tidal forces which can suppress the former \citep{Liu2015} and treatments for eccentric mass-transfer \citep{HamersDosopoulou2019}, which is left for future work. For massive stars our approximation is well justified in many cases because their lifetime of a few Myr is often shorter than the timescale $t_{\rm LK}$ of three-body effects. Here, we adopt a conservative approach and discard from our analysis any merger with a short delay time $t_{\rm d}<t_{\rm th}$, since its progenitor binary might instead have undergone a stellar merger before compact-object formation. As a default, we adopt a threshold $t_{\rm th}=5\,\rm Myr$, and have verified that our inferred distributions do not depend sensitively on this value. iii) Our simulations account for the possibility that the tertiary companion also forms a black hole or neutron star, in which case we update the parameters of the outer orbit and check for orbital disruption. However, if it is a low-mass star we treat it effectively as a non-evolving point mass in our three-body dynamics integration after inner binary black hole formation. This neglects the effect of subsequent wind-mass loss of the tertiary companion or white dwarf formation. We expect this effect to be negligible on a population level due to the weakness of winds from low-mass stars and the long lifetime of low-mass stars; iv) We ignore the impact of higher-order companions in hierarchical quadruples, quintuples, etc. While their evolution may give rise to interesting higher-order dynamical effects, most of them effectively evolve like a hierarchical triple to leading order \citep{Vynatheya2022}. 

\begin{appendices}

\section{Model variations}\label{sec:models}

\begin{figure}
    \centering
    \includegraphics[width=0.7\linewidth]{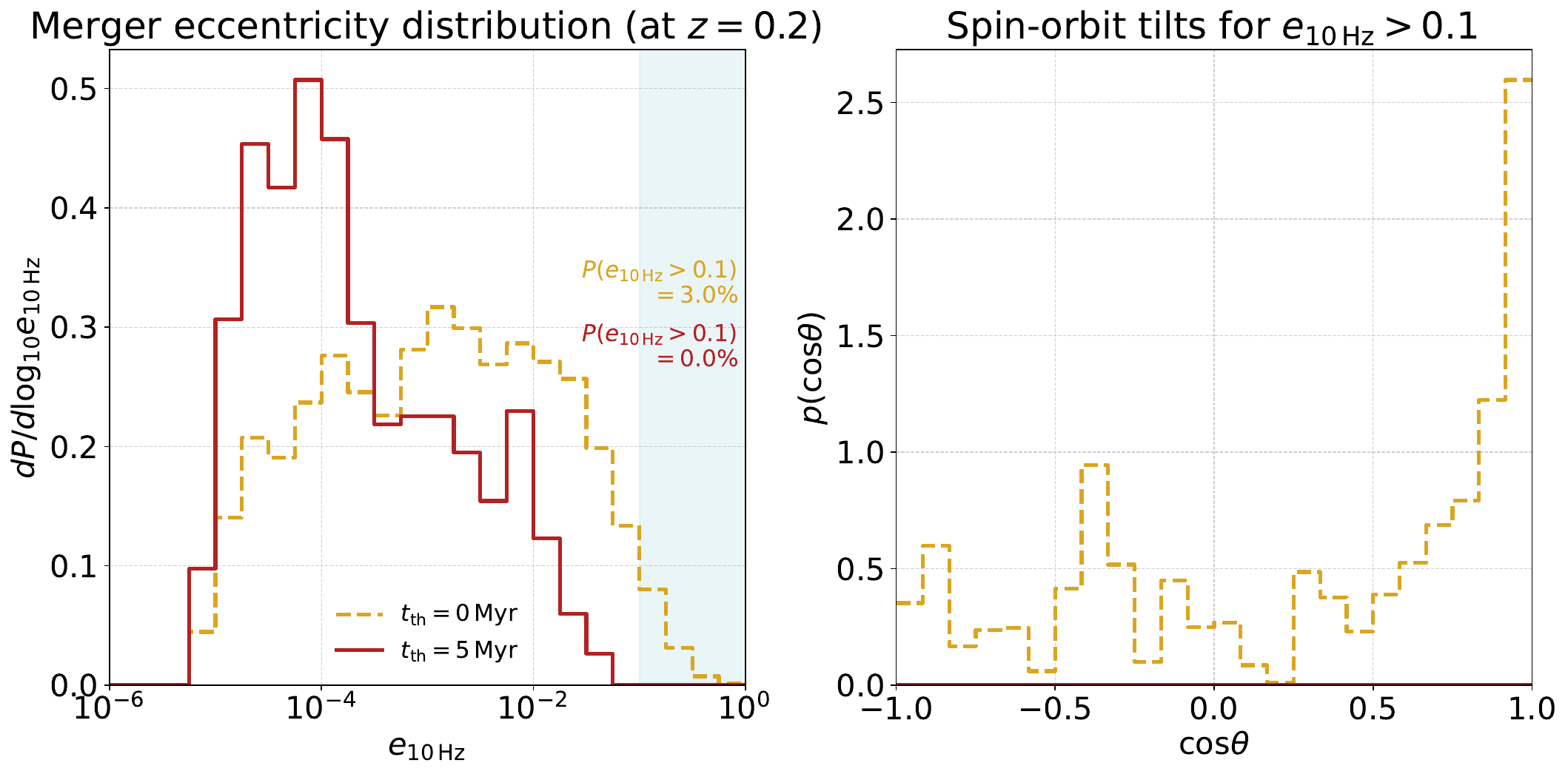}
    \caption{Left: Distribution of inner orbital eccentricities of mergers at $z=0.2$. They are recorded at a reference frequency of $10\,\rm Hz$, i.e., $f_{\rm GW}(e_1=e_{10\,\rm Hz})=10\,\rm Hz$ in Eq.~\eqref{eq:fGW}. Different colours indicate different delay time thresholds below which we discard mergers (see Sec.~\ref{sec4}). Gold corresponds to no delay time cut ($t_{\rm th}=0$); red assumes the typical lifetime of massive stars ($t_{\rm th}=5\,\rm Myr$). The resulting fraction of mergers with very large, potentially detectable eccentricities ($e_{10\,\rm Hz}>0.1$) ranges from a few percent to zero, respectively. Right: Spin-orbit tilt distribution of mergers with $e_{10\,\rm Hz}>0.1$.}
    \label{fig:eccentricity}
\end{figure}

\begin{figure}
    \centering
    \includegraphics[width=0.7\linewidth]{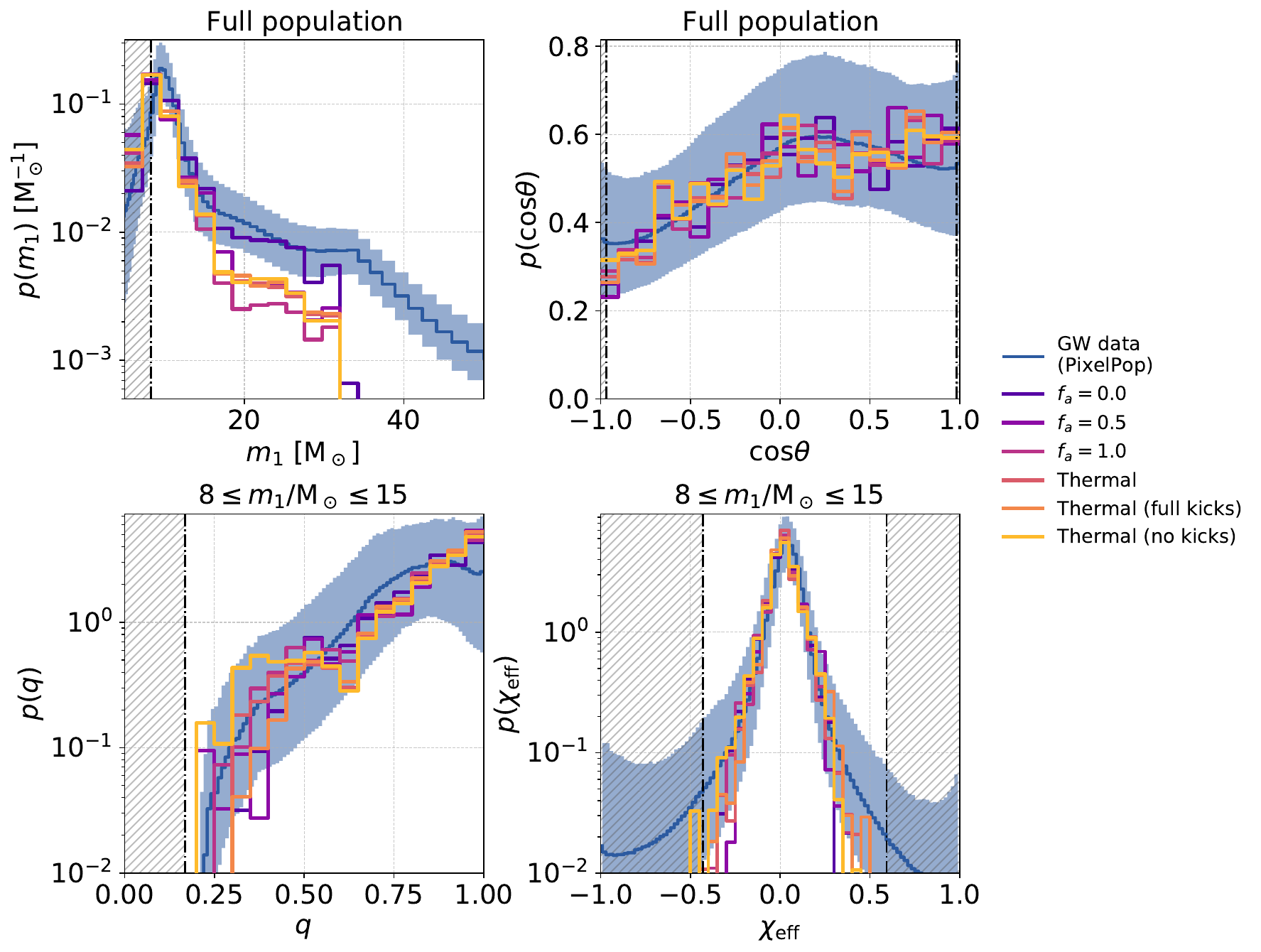}
    \caption{Comparison of models with different mass-transfer efficiencies and black-hole natal-kick prescriptions. Top row: Distributions of primary black-hole mass (left) and spin-orbit tilt (right) for the full merger population. Bottom row: Mass-ratio (left) and effective-spin (right) distributions for systems in the primary-mass range $8\,\Msun \leq m_1 \leq 15\,\Msun$. For the effective-spin distribution, we adopt a fiducial black-hole spin-magnitude distribution with $(\mu_\chi,\sigma_\chi)=(0.1,0.2)$. All distributions are evaluated at redshift $z=0.2$.}
    \label{fig:model-comp}
\end{figure}

Here, we present results for several model variations exploring the impact of two highly uncertain aspects of binary evolution: the mass-transfer efficiency and natal black-hole kicks. We consider four prescriptions for the mass-transfer efficiency onto the stellar companion: fully non-conservative mass transfer (\(f_a=0.0\)); a fixed efficiency of \(f_a=0.5\), for which half of the transferred mass is accreted by the companion and half is lost from the system; fully conservative mass transfer (\(f_a=1.0\)); and the \texttt{COMPAS} default prescription, in which the accretion efficiency is determined by the thermal timescales of the donor and accretor \citep{Hurley2002}. We additionally test three prescriptions for natal kicks: no natal kicks, the standard fallback-reduced kicks, and full natal kicks without fallback reduction. The underlying kick magnitudes are drawn following the default \texttt{COMPAS} prescription based on Ref.~\cite{MandelMuller2020}.

As shown in Fig.~\ref{fig:model-comp}, despite the substantial differences between these assumptions, neither the mass-transfer efficiency nor the natal-kick prescription has a significant impact on the overall properties of the merging black-hole population. In particular, the merger rates, mass-ratio distributions, and black-hole spin distributions remain largely unchanged. The most noticeable difference is in the contribution of systems with \(m_1 \gtrsim 15\,\Msun\), which is enhanced for non-conservative mass transfer compared to other models.




\end{appendices}


\section*{Acknowledgements}
We thank Fabio Antonini, Lieke van Son, and Selma de Mink for helpful input. This work was performed in part at the Aspen Center for Physics, which is supported by the National Science Foundation grant PHY-2210452 and a grant from the Simons Foundation (1161654, Troyer). 
A.O.~acknowledges support from the Polish National Agency for Academic Exchange (NAWA) through the Polish Returns Programme, grant no. BPN/PPO/2025/1/00007/U/00001, and from the National Science Centre, Poland (NCN), through OPUS 30 grant no. 2025/59/B/ST9/01180. We acknowledge project support from the Max Planck Computing and Data Facility.

\section*{Author contributions}
J.S.~conceived the study, set up the simulations, and performed the numerical analysis of the outcomes. J.S.~and A.O.~contributed equally to all other parts of this work, including the choice of astrophysical models, interpretations of the findings, and manuscript writing.  

\bibliography{sn-bibliography}

@ARTICLE{Olejak2024,
       author = {{Olejak}, Aleksandra and {Klencki}, Jakub and {Xu}, Xiao-Tian and {Wang}, Chen and {Belczynski}, Krzysztof and {Lasota}, Jean-Pierre},
        title = "{Unequal-mass highly spinning binary black hole mergers in the stable mass transfer formation channel}",
      journal = {\aap},
         year = 2024,
        month = sep,
       volume = {689},
          eid = {A305},
        pages = {A305},
          doi = {10.1051/0004-6361/202450480},
archivePrefix = {arXiv},
       eprint = {2404.12426},
 primaryClass = {astro-ph.HE},
       adsurl = {https://ui.adsabs.harvard.edu/abs/2024A&A...689A.305O}
}

@ARTICLE{Gerosa2018,
       author = {{Gerosa}, Davide and {Berti}, Emanuele and {O'Shaughnessy}, Richard and {Belczynski}, Krzysztof and {Kesden}, Michael and {Wysocki}, Daniel and {Gladysz}, Wojciech},
        title = "{Spin orientations of merging black holes formed from the evolution of stellar binaries}",
      journal = {\prd},
         year = 2018,
        month = oct,
       volume = {98},
       number = {8},
          eid = {084036},
        pages = {084036},
          doi = {10.1103/PhysRevD.98.084036},
archivePrefix = {arXiv},
       eprint = {1808.02491},
 primaryClass = {astro-ph.HE},
       adsurl = {https://ui.adsabs.harvard.edu/abs/2018PhRvD..98h4036G}
}

@ARTICLE{Dorozsmai2025,
       author = {{Dorozsmai}, Andris and {Romero-Shaw}, Isobel M. and {Vijaykumar}, Aditya and {Toonen}, Silvia and {Antonini}, Fabio and {Kremer}, Kyle and {Zevin}, Michael and {Grishin}, Evgeni},
        title = "{Hierarchical Triples vs. Globular Clusters: Binary black hole merger eccentricity distributions compete and evolve with redshift}",
      journal = {\mnras},
         year = 2025,
        month = nov,
          doi = {10.1093/mnras/staf1938},
archivePrefix = {arXiv},
       eprint = {2507.23212},
 primaryClass = {astro-ph.GA},
       adsurl = {https://ui.adsabs.harvard.edu/abs/2025MNRAS.tmp.1834D}
}

@article{Hurley2002,
	adsurl = {https://ui.adsabs.harvard.edu/abs/2002MNRAS.329..897H},
	archiveprefix = {arXiv},
	author = {{Hurley}, Jarrod R. and {Tout}, Christopher A. and {Pols}, Onno R.},
	doi = {10.1046/j.1365-8711.2002.05038.x},
	eprint = {astro-ph/0201220},
	journal = {Mon. Not. R. Astron. Soc.},
	month = feb,
	number = {4},
	pages = {897-928},
	primaryclass = {astro-ph},
	title = {{Evolution of binary stars and the effect of tides on binary populations}},
	volume = {329},
	year = 2002}

@article{Rodriguez2018,
	adsurl = {https://ui.adsabs.harvard.edu/abs/2018ApJ...863....7R},
	archiveprefix = {arXiv},
	author = {{Rodriguez}, Carl L. and {Antonini}, Fabio},
	doi = {10.3847/1538-4357/aacea4},
	eid = {7},
	eprint = {1805.08212},
	journal = {Astrophys. J.},
	month = aug,
	number = {1},
	pages = {7},
	primaryclass = {astro-ph.HE},
	title = {{A Triple Origin for the Heavy and Low-spin Binary Black Holes Detected by LIGO/VIRGO}},
	volume = {863},
	year = 2018}

@article{Lidov1962,
	adsurl = {https://ui.adsabs.harvard.edu/abs/1962P&SS....9..719L},
	author = {{Lidov}, M.~L.},
	doi = {10.1016/0032-0633(62)90129-0},
	journal = {Planet. Space Sci.},
	month = oct,
	number = {10},
	pages = {719-759},
	title = {{The evolution of orbits of artificial satellites of planets under the action of gravitational perturbations of external bodies}},
	volume = {9},
	year = 1962}

@article{Kozai1962,
	adsurl = {https://ui.adsabs.harvard.edu/abs/1962AJ.....67..591K},
	author = {{Kozai}, Yoshihide},
	doi = {10.1086/108790},
	journal = {Astron. J.},
	month = nov,
	pages = {591-598},
	title = {{Secular perturbations of asteroids with high inclination and eccentricity}},
	volume = {67},
	year = 1962}

@article{GW150914,
	adsurl = {https://ui.adsabs.harvard.edu/abs/2016PhRvL.116f1102A},
	archiveprefix = {arXiv},
	author = {{Abbott}, B.~P. and others},
	doi = {10.1103/PhysRevLett.116.061102},
	eid = {061102},
	eprint = {1602.03837},
	journal = {Phys. Rev. Lett.},
	month = feb,
	number = {6},
	pages = {061102},
	primaryclass = {gr-qc},
	title = {{Observation of Gravitational Waves from a Binary Black Hole Merger}},
	volume = {116},
	year = 2016}

@article{Neijssel2019,
	adsurl = {https://ui.adsabs.harvard.edu/abs/2019MNRAS.490.3740N},
	archiveprefix = {arXiv},
	author = {{Neijssel}, Coenraad J. and {Vigna-G{\'o}mez}, Alejandro and {Stevenson}, Simon and {Barrett}, Jim W. and {Gaebel}, Sebastian M. and {Broekgaarden}, Floor S. and {de Mink}, Selma E. and {Sz{\'e}csi}, Dorottya and {Vinciguerra}, Serena and {Mandel}, Ilya},
	doi = {10.1093/mnras/stz2840},
	eprint = {1906.08136},
	journal = {Mon. Not. R. Astron. Soc.},
	month = dec,
	number = {3},
	pages = {3740-3759},
	primaryclass = {astro-ph.SR},
	title = {{The effect of the metallicity-specific star formation history on double compact object mergers}},
	volume = {490},
	year = 2019}

@ARTICLE{2025A&A...699A.272V,
       author = {{Vigna-G{\'o}mez}, A. and {Grishin}, E. and {Stegmann}, J. and {Olejak}, A. and {Popa}, S.~A. and {Liu}, B. and {Rajamuthukumar}, A.~S. and {van Son}, L.~A.~C. and {Bobrick}, A. and {Dorozsmai}, A.},
        title = "{Prompt stellar and binary black hole mergers in tight triples: Insights from chemically homogeneous evolution}",
      journal = {\aap},
         year = 2025,
        month = jul,
       volume = {699},
          eid = {A272},
        pages = {A272},
          doi = {10.1051/0004-6361/202554680},
archivePrefix = {arXiv},
       eprint = {2503.17006},
 primaryClass = {astro-ph.SR},
       adsurl = {https://ui.adsabs.harvard.edu/abs/2025A&A...699A.272V}
}

@ARTICLE{Rodriguez2016,
       author = {{Rodriguez}, Carl L. and {Zevin}, Michael and {Pankow}, Chris and {Kalogera}, Vasilliki and {Rasio}, Frederic A.},
        title = "{Illuminating Black Hole Binary Formation Channels with Spins in Advanced LIGO}",
      journal = {\apjl},
         year = 2016,
        month = nov,
       volume = {832},
       number = {1},
          eid = {L2},
        pages = {L2},
          doi = {10.3847/2041-8205/832/1/L2},
archivePrefix = {arXiv},
       eprint = {1609.05916},
 primaryClass = {astro-ph.HE},
       adsurl = {https://ui.adsabs.harvard.edu/abs/2016ApJ...832L...2R}
}

@article{Liu2018,
	adsurl = {https://ui.adsabs.harvard.edu/abs/2018ApJ...863...68L},
	archiveprefix = {arXiv},
	author = {{Liu}, Bin and {Lai}, Dong},
	doi = {10.3847/1538-4357/aad09f},
	eid = {68},
	eprint = {1805.03202},
	journal = {Astrophys. J.},
	month = aug,
	number = {1},
	pages = {68},
	primaryclass = {astro-ph.HE},
	title = {{Black Hole and Neutron Star Binary Mergers in Triple Systems: Merger Fraction and Spin-Orbit Misalignment}},
	volume = {863},
	year = 2018}

@article{Antonini2018,
	adsurl = {https://ui.adsabs.harvard.edu/abs/2018MNRAS.480L..58A},
	archiveprefix = {arXiv},
	author = {{Antonini}, Fabio and {Rodriguez}, Carl L. and {Petrovich}, Cristobal and {Fischer}, Caitlin L.},
	doi = {10.1093/mnrasl/sly126},
	eprint = {1711.07142},
	journal = {Mon. Not. R. Astron. Soc.},
	month = oct,
	number = {1},
	pages = {L58-L62},
	primaryclass = {astro-ph.HE},
	title = {{Precessional dynamics of black hole triples: binary mergers with near-zero effective spin}},
	volume = {480},
	year = 2018}

@article{Schmidt2015,
  author        = {Schmidt, Patricia and Ohme, Frank and Hannam, Mark},
  title         = {Towards Models of Gravitational Waveforms from Generic
                   Binaries. II. Modelling Precession Effects with a Single
                   Effective Precession Parameter},
  journal       = {Physical Review D},
  volume        = {91},
  number        = {2},
  pages         = {024043},
  year          = {2015},
  doi           = {10.1103/PhysRevD.91.024043},
  eprint        = {1408.1810},
  archiveprefix = {arXiv},
  primaryclass  = {gr-qc}
}

@article{Antonini2017,
	adsurl = {https://ui.adsabs.harvard.edu/abs/2017ApJ...841...77A},
	archiveprefix = {arXiv},
	author = {{Antonini}, Fabio and {Toonen}, Silvia and {Hamers}, Adrian S.},
	doi = {10.3847/1538-4357/aa6f5e},
	eid = {77},
	eprint = {1703.06614},
	journal = {Astrophys. J.},
	month = jun,
	number = {2},
	pages = {77},
	primaryclass = {astro-ph.GA},
	title = {{Binary Black Hole Mergers from Field Triples: Properties, Rates, and the Impact of Stellar Evolution}},
	volume = {841},
	year = 2017}

@article{Silsbee2017,
	adsurl = {https://ui.adsabs.harvard.edu/abs/2017ApJ...836...39S},
	archiveprefix = {arXiv},
	author = {{Silsbee}, Kedron and {Tremaine}, Scott},
	doi = {10.3847/1538-4357/aa5729},
	eid = {39},
	eprint = {1608.07642},
	journal = {Astrophys. J.},
	month = feb,
	number = {1},
	pages = {39},
	primaryclass = {astro-ph.HE},
	title = {{Lidov-Kozai Cycles with Gravitational Radiation: Merging Black Holes in Isolated Triple Systems}},
	volume = {836},
	year = 2017}

@ARTICLE{AntoniniGieles2023,
       author = {{Antonini}, Fabio and {Gieles}, Mark and {Dosopoulou}, Fani and {Chattopadhyay}, Debatri},
        title = "{Coalescing black hole binaries from globular clusters: mass distributions and comparison to gravitational wave data from GWTC-3}",
      journal = {\mnras},
         year = 2023,
        month = jun,
       volume = {522},
       number = {1},
        pages = {466-476},
          doi = {10.1093/mnras/stad972},
archivePrefix = {arXiv},
       eprint = {2208.01081},
 primaryClass = {astro-ph.HE},
       adsurl = {https://ui.adsabs.harvard.edu/abs/2023MNRAS.522..466A}
}

@ARTICLE{Stegmann2025,
       author = {{Stegmann}, Jakob and {Klencki}, Jakub},
        title = "{Orbital Eccentricity and Spin{\textendash}Orbit Misalignment Are Evidence that Neutron Star{\textendash}Black Hole Mergers Form through Triple Star Evolution}",
      journal = {\apjl},
         year = 2025,
        month = oct,
       volume = {991},
       number = {2},
          eid = {L54},
        pages = {L54},
          doi = {10.3847/2041-8213/ae055b},
       adsurl = {https://ui.adsabs.harvard.edu/abs/2025ApJ...991L..54S}
}

@article{MandelBroekgaarden2022,
	adsurl = {https://ui.adsabs.harvard.edu/abs/2022LRR....25....1M},
	archiveprefix = {arXiv},
	author = {{Mandel}, Ilya and {Broekgaarden}, Floor S.},
	doi = {10.1007/s41114-021-00034-3},
	eid = {1},
	eprint = {2107.14239},
	journal = {Living Rev. Relativ.},
	month = dec,
	number = {1},
	pages = {1},
	primaryclass = {astro-ph.HE},
	title = {{Rates of compact object coalescences}},
	volume = {25},
	year = 2022}

@article{El-Badry2018,
	adsurl = {https://ui.adsabs.harvard.edu/abs/2018MNRAS.480.4884E},
	archiveprefix = {arXiv},
	author = {{El-Badry}, Kareem and {Rix}, Hans-Walter},
	doi = {10.1093/mnras/sty2186},
	eprint = {1807.06011},
	journal = {Mon. Not. R. Astron. Soc.},
	month = nov,
	number = {4},
	pages = {4884-4902},
	primaryclass = {astro-ph.SR},
	title = {{Imprints of white dwarf recoil in the separation distribution of Gaia wide binaries}},
	volume = {480},
	year = 2018}

@article{Sana2012,
	adsurl = {https://ui.adsabs.harvard.edu/abs/2012Sci...337..444S},
	archiveprefix = {arXiv},
	author = {{Sana}, H. and {de Mink}, S.~E. and {de Koter}, A. and {Langer}, N. and {Evans}, C.~J. and {Gieles}, M. and {Gosset}, E. and {Izzard}, R.~G. and {Le Bouquin}, J. -B. and {Schneider}, F.~R.~N.},
	doi = {10.1126/science.1223344},
	eprint = {1207.6397},
	journal = {Science},
	month = jul,
	number = {6093},
	pages = {444},
	primaryclass = {astro-ph.SR},
	title = {{Binary Interaction Dominates the Evolution of Massive Stars}},
	volume = {337},
	year = 2012}

@article{Mandel2016,
	adsurl = {https://ui.adsabs.harvard.edu/abs/2016MNRAS.456..578M},
	archiveprefix = {arXiv},
	author = {{Mandel}, Ilya},
	doi = {10.1093/mnras/stv2733},
	eprint = {1510.03871},
	journal = {Mon. Not. R. Astron. Soc.},
	month = feb,
	number = {1},
	pages = {578-581},
	primaryclass = {astro-ph.HE},
	title = {{Estimates of black hole natal kick velocities from observations of low-mass X-ray binaries}},
	volume = {456},
	year = 2016}

@article{Heger2003,
	adsurl = {https://ui.adsabs.harvard.edu/abs/2003ApJ...591..288H},
	archiveprefix = {arXiv},
	author = {{Heger}, A. and {Fryer}, C.~L. and {Woosley}, S.~E. and {Langer}, N. and {Hartmann}, D.~H.},
	doi = {10.1086/375341},
	eprint = {astro-ph/0212469},
	journal = {Astrophys. J.},
	month = jul,
	number = {1},
	pages = {288-300},
	primaryclass = {astro-ph},
	title = {{How Massive Single Stars End Their Life}},
	volume = {591},
	year = 2003}

@article{vanSon2022,
	adsurl = {https://ui.adsabs.harvard.edu/abs/2022ApJ...940..184V},
	archiveprefix = {arXiv},
	author = {{van Son}, L.~A.~C. and {de Mink}, S.~E. and {Renzo}, M. and {Justham}, S. and {Zapartas}, E. and {Breivik}, K. and {Callister}, T. and {Farr}, W.~M. and {Conroy}, C.},
	doi = {10.3847/1538-4357/ac9b0a},
	eid = {184},
	eprint = {2209.13609},
	journal = {Astrophys. J.},
	month = dec,
	number = {2},
	pages = {184},
	primaryclass = {astro-ph.HE},
	title = {{No Peaks without Valleys: The Stable Mass Transfer Channel for Gravitational-wave Sources in Light of the Neutron Star-Black Hole Mass Gap}},
	volume = {940},
	year = 2022}

@article{Olejak2021,
	adsurl = {https://ui.adsabs.harvard.edu/abs/2021A&A...651A.100O},
	archiveprefix = {arXiv},
	author = {{Olejak}, A. and {Belczynski}, K. and {Ivanova}, N.},
	doi = {10.1051/0004-6361/202140520},
	eid = {A100},
	eprint = {2102.05649},
	journal = {Astron. Astrophys.},
	month = jul,
	pages = {A100},
	primaryclass = {astro-ph.HE},
	title = {{Impact of common envelope development criteria on the formation of LIGO/Virgo sources}},
	volume = {651},
	year = 2021}

@article{Stegmann2022b,
	adsurl = {https://ui.adsabs.harvard.edu/abs/2022PhRvD.106b3014S},
	archiveprefix = {arXiv},
	author = {{Stegmann}, Jakob and {Antonini}, Fabio and {Schneider}, Fabian R.~N. and {Tiwari}, Vaibhav and {Chattopadhyay}, Debatri},
	doi = {10.1103/PhysRevD.106.023014},
	eid = {023014},
	eprint = {2203.16544},
	journal = {Phys. Rev. D},
	month = jul,
	number = {2},
	pages = {023014},
	primaryclass = {astro-ph.GA},
	title = {{Binary black hole mergers from merged stars in the Galactic field}},
	volume = {106},
	year = 2022}

@article{Schneider2023,
	adsurl = {https://ui.adsabs.harvard.edu/abs/2023ApJ...950L...9S},
	archiveprefix = {arXiv},
	author = {{Schneider}, Fabian R.~N. and {Podsiadlowski}, Philipp and {Laplace}, Eva},
	doi = {10.3847/2041-8213/acd77a},
	eid = {L9},
	eprint = {2305.02380},
	journal = {Astrophys. J. Lett.},
	month = jun,
	number = {2},
	pages = {L9},
	primaryclass = {astro-ph.HE},
	title = {{Bimodal Black Hole Mass Distribution and Chirp Masses of Binary Black Hole Mergers}},
	volume = {950},
	year = 2023}

@article{Antonini2014,
	adsurl = {https://ui.adsabs.harvard.edu/abs/2014ApJ...781...45A},
	archiveprefix = {arXiv},
	author = {{Antonini}, Fabio and {Murray}, Norman and {Mikkola}, Seppo},
	doi = {10.1088/0004-637X/781/1/45},
	eid = {45},
	eprint = {1308.3674},
	journal = {Astrophys. J.},
	month = jan,
	number = {1},
	pages = {45},
	primaryclass = {astro-ph.HE},
	title = {{Black Hole Triple Dynamics: A Breakdown of the Orbit Average Approximation and Implications for Gravitational Wave Detections}},
	volume = {781},
	year = 2014}

@article{Grishin2018,
	adsurl = {https://ui.adsabs.harvard.edu/abs/2018MNRAS.481.4907G},
	archiveprefix = {arXiv},
	author = {{Grishin}, Evgeni and {Perets}, Hagai B. and {Fragione}, Giacomo},
	doi = {10.1093/mnras/sty2477},
	eprint = {1808.02030},
	journal = {Mon. Not. R. Astron. Soc.},
	month = dec,
	number = {4},
	pages = {4907-4923},
	primaryclass = {astro-ph.EP},
	title = {{Quasi-secular evolution of mildly hierarchical triple systems: analytics and applications for GW sources and hot Jupiters}},
	volume = {481},
	year = 2018}

@article{Naoz2013,
	adsurl = {https://ui.adsabs.harvard.edu/abs/2013MNRAS.431.2155N},
	archiveprefix = {arXiv},
	author = {{Naoz}, Smadar and {Farr}, Will M. and {Lithwick}, Yoram and {Rasio}, Frederic A. and {Teyssandier}, Jean},
	doi = {10.1093/mnras/stt302},
	eprint = {1107.2414},
	journal = {Mon. Not. R. Astron. Soc.},
	month = may,
	number = {3},
	pages = {2155-2171},
	primaryclass = {astro-ph.EP},
	title = {{Secular dynamics in hierarchical three-body systems}},
	volume = {431},
	year = 2013}

@article{Stegmann2022,
	adsurl = {https://ui.adsabs.harvard.edu/abs/2022MNRAS.516.1406S},
	archiveprefix = {arXiv},
	author = {{Stegmann}, Jakob and {Antonini}, Fabio and {Moe}, Maxwell},
	doi = {10.1093/mnras/stac2192},
	eprint = {2112.10786},
	journal = {Mon. Not. R. Astron. Soc.},
	month = oct,
	number = {1},
	pages = {1406-1427},
	primaryclass = {astro-ph.SR},
	title = {{Evolution of massive stellar triples and implications for compact object binary formation}},
	volume = {516},
	year = 2022}

@article{Wen2003,
	adsurl = {https://ui.adsabs.harvard.edu/abs/2003ApJ...598..419W},
	archiveprefix = {arXiv},
	author = {{Wen}, Linqing},
	doi = {10.1086/378794},
	eprint = {astro-ph/0211492},
	journal = {Astrophys. J.},
	month = nov,
	number = {1},
	pages = {419-430},
	primaryclass = {astro-ph},
	title = {{On the Eccentricity Distribution of Coalescing Black Hole Binaries Driven by the Kozai Mechanism in Globular Clusters}},
	volume = {598},
	year = 2003}

@ARTICLE{Farag2022,
       author = {{Farag}, Ebraheem and {Renzo}, Mathieu and {Farmer}, Robert and {Chidester}, Morgan T. and {Timmes}, F.~X.},
        title = "{Resolving the Peak of the Black Hole Mass Spectrum}",
      journal = {\apj},
         year = 2022,
        month = oct,
       volume = {937},
       number = {2},
          eid = {112},
        pages = {112},
          doi = {10.3847/1538-4357/ac8b83},
archivePrefix = {arXiv},
       eprint = {2208.09624},
 primaryClass = {astro-ph.HE},
       adsurl = {https://ui.adsabs.harvard.edu/abs/2022ApJ...937..112F}
}

@ARTICLE{Farmer2019,
       author = {{Farmer}, R. and {Renzo}, M. and {de Mink}, S.~E. and {Marchant}, P. and {Justham}, S.},
        title = "{Mind the Gap: The Location of the Lower Edge of the Pair-instability Supernova Black Hole Mass Gap}",
      journal = {\apj},
         year = 2019,
        month = dec,
       volume = {887},
       number = {1},
          eid = {53},
        pages = {53},
          doi = {10.3847/1538-4357/ab518b},
archivePrefix = {arXiv},
       eprint = {1910.12874},
 primaryClass = {astro-ph.SR},
       adsurl = {https://ui.adsabs.harvard.edu/abs/2019ApJ...887...53F}
}

@article{Barnes2018,
  author   = {Barnes, Luke A.},
  title    = {Fine-tuning in the context of {Bayesian} theory testing},
  journal  = {European Journal for Philosophy of Science},
  year     = {2018},
  month    = may,
  volume   = {8},
  number   = {2},
  pages    = {253--269},
  issn     = {1879-4920},
  doi      = {10.1007/s13194-017-0184-2},
  url      = {https://doi.org/10.1007/s13194-017-0184-2}
}

@ARTICLE{Bruenech2026,
       author = {{Bruenech}, C.~W. and {Toonen}, S. and {Boekholt}, T. and {Dorozsmai}, A.},
        title = "{Triple-induced mergers of black hole binaries: A comprehensive look at the role of stellar evolution, dynamical stability, and spin evolution}",
      journal = {arXiv e-prints},
         year = 2026,
        month = aug,
          eid = {arXiv:2608.25538},
        pages = {arXiv:2608.25538},
          doi = {10.48550/arXiv.2608.25538},
archivePrefix = {arXiv},
       eprint = {2608.25538},
 primaryClass = {astro-ph.SR},
       adsurl = {https://ui.adsabs.harvard.edu/abs/2026arXiv260825538B}
}

@misc{gallegosgarcia2026massratiodistributionlowmassbinary,
      title={The Mass-Ratio Distribution of the Low-Mass Binary Black Hole Subpopulation: A Natural Outcome of Isolated Binary Evolution}, 
      author={Monica Gallegos-Garcia and Anarya Ray and Vicky Kalogera and Max Briel and Michael Zevin and Abhishek Chattaraj and Zepei Xing and Jeff J. Andrews and Seth Gossage and Philipp M. Srivastava},
      year={2026},
      eprint={2609.01784},
      archivePrefix={arXiv},
      primaryClass={astro-ph.HE},
      url={https://arxiv.org/abs/2609.01784}, 
}

@misc{rodriguez2019kozai,
  author       = {Rodriguez, Carl},
  title        = {{Carl's Kozai Code: C++ Version}},
  year         = {2019},
  howpublished = {\url{https://github.com/carlrodriguez/kozai/tree/master/C++_Version}},
  note         = {GitHub repository, accessed 2026-09-03}
}

@article{LuNaoz,
    author = {Lu, Cicero X and Naoz, Smadar},
    title = {Supernovae kicks in hierarchical triple systems},
    journal = {Monthly Notices of the Royal Astronomical Society},
    volume = {484},
    number = {2},
    pages = {1506-1525},
    year = {2019},
    month = {04},
    issn = {0035-8711},
    doi = {10.1093/mnras/stz036},
    url = {https://doi.org/10.1093/mnras/stz036},
    eprint = {https://academic.oup.com/mnras/article-pdf/484/2/1506/27583150/stz036.pdf},
}

@article{PhysRevLett.120.151101,
  title = {Post-Newtonian Dynamics in Dense Star Clusters: Highly Eccentric, Highly Spinning, and Repeated Binary Black Hole Mergers},
  author = {Rodriguez, Carl L. and Amaro-Seoane, Pau and Chatterjee, Sourav and Rasio, Frederic A.},
  journal = {Phys. Rev. Lett.},
  volume = {120},
  issue = {15},
  pages = {151101},
  numpages = {7},
  year = {2018},
  month = {Apr},
  publisher = {American Physical Society},
  doi = {10.1103/PhysRevLett.120.151101},
  url = {https://link.aps.org/doi/10.1103/PhysRevLett.120.151101}
}

@ARTICLE{Farr2018,
       author = {{Farr}, Ben and {Holz}, Daniel E. and {Farr}, Will M.},
        title = "{Using Spin to Understand the Formation of LIGO and Virgo{\textquoteright}s Black Holes}",
      journal = {\apjl},
         year = 2018,
        month = feb,
       volume = {854},
       number = {1},
          eid = {L9},
        pages = {L9},
          doi = {10.3847/2041-8213/aaaa64},
archivePrefix = {arXiv},
       eprint = {1709.07896},
 primaryClass = {astro-ph.HE},
       adsurl = {https://ui.adsabs.harvard.edu/abs/2018ApJ...854L...9F}
}

@ARTICLE{2025ApJ...987..164B,
       author = {{Burrows}, Adam and {Wang}, Tianshu and {Vartanyan}, David},
        title = "{Channels of Stellar-mass Black Hole Formation}",
      journal = {\apj},
         year = 2025,
        month = jul,
       volume = {987},
       number = {2},
          eid = {164},
        pages = {164},
          doi = {10.3847/1538-4357/addd04},
archivePrefix = {arXiv},
       eprint = {2412.07831},
 primaryClass = {astro-ph.SR},
       adsurl = {https://ui.adsabs.harvard.edu/abs/2025ApJ...987..164B}
}

@ARTICLE{BaibhavKalogera2024,
       author = {{Baibhav}, Vishal and {Kalogera}, Vicky},
        title = "{Revising the Spin and Kick Connection in Isolated Binary Black Holes}",
      journal = {arXiv e-prints},
         year = 2024,
        month = dec,
          eid = {arXiv:2412.03461},
        pages = {arXiv:2412.03461},
          doi = {10.48550/arXiv.2412.03461},
archivePrefix = {arXiv},
       eprint = {2412.03461},
 primaryClass = {astro-ph.HE},
       adsurl = {https://ui.adsabs.harvard.edu/abs/2024arXiv241203461B}
}

@misc{whitaker2026longperiodstellarmassblack,
      title={A Long Period Stellar-Mass Black Hole Binary in $\omega$ Centauri}, 
      author={Matthew Whitaker and Evan Kerr and Anil Seth and Maximilian Häberle and Jay Strader and Jay Anderson and Andrea Bellini and Callie Clontz and Zack Freeman and Massimo Griggio and Sebastian Kamann and Mattia Libralato and Nadine Neumayer and Elena González Prieto and Carl L. Rodriguez and Sara Saracino and Peter Smith and Glenn van de Ven and Zixian Wang},
      year={2026},
      eprint={2606.18350},
      archivePrefix={arXiv},
      primaryClass={astro-ph.GA},
      url={https://arxiv.org/abs/2606.18350}, 
}

@ARTICLE{2024PhRvL.132s1403V,
       author = {{Vigna-G{\'o}mez}, Alejandro and {Willcox}, Reinhold and {Tamborra}, Irene and {Mandel}, Ilya and {Renzo}, Mathieu and {Wagg}, Tom and {Janka}, Hans-Thomas and {Kresse}, Daniel and {Bodensteiner}, Julia and {Shenar}, Tomer and {Tauris}, Thomas M.},
        title = "{Constraints on Neutrino Natal Kicks from Black-Hole Binary VFTS 243}",
      journal = {\prl},
         year = 2024,
        month = may,
       volume = {132},
       number = {19},
          eid = {191403},
        pages = {191403},
          doi = {10.1103/PhysRevLett.132.191403},
archivePrefix = {arXiv},
       eprint = {2310.01509},
 primaryClass = {astro-ph.HE},
       adsurl = {https://ui.adsabs.harvard.edu/abs/2024PhRvL.132s1403V}
}

@ARTICLE{2024Ap&SS.369...80J,
       author = {{Janka}, Hans-Thomas and {Kresse}, Daniel},
        title = "{Interplay between neutrino kicks and hydrodynamic kicks of neutron stars and black holes}",
      journal = {\apss},
         year = 2024,
        month = aug,
       volume = {369},
       number = {8},
          eid = {80},
        pages = {80},
          doi = {10.1007/s10509-024-04343-1},
archivePrefix = {arXiv},
       eprint = {2401.13817},
 primaryClass = {astro-ph.HE},
       adsurl = {https://ui.adsabs.harvard.edu/abs/2024Ap&SS.369...80J}
}

@ARTICLE{vanSon2023,
       author = {{van Son}, L.~A.~C. and {de Mink}, S.~E. and {Chru{\'s}li{\'n}ska}, M. and {Conroy}, C. and {Pakmor}, R. and {Hernquist}, L.},
        title = "{The Locations of Features in the Mass Distribution of Merging Binary Black Holes Are Robust against Uncertainties in the Metallicity-dependent Cosmic Star Formation History}",
      journal = {\apj},
         year = 2023,
        month = may,
       volume = {948},
       number = {2},
          eid = {105},
        pages = {105},
          doi = {10.3847/1538-4357/acbf51},
archivePrefix = {arXiv},
       eprint = {2209.03385},
 primaryClass = {astro-ph.GA},
       adsurl = {https://ui.adsabs.harvard.edu/abs/2023ApJ...948..105V}
}

@INCOLLECTION{Mapellibook2021,
       author = {{Mapelli}, Michela},
        title = "{Formation Channels of Single and Binary Stellar-Mass Black Holes}",
    booktitle = {Handbook of Gravitational Wave Astronomy},
         year = 2021,
       editor = {{Bambi}, Cosimo and {Katsanevas}, Stavros and {Kokkotas}, Konstantinos D.},
          eid = {16},
        pages = {16},
          doi = {10.1007/978-981-15-4702-7_16-1},
       adsurl = {https://ui.adsabs.harvard.edu/abs/2021hgwa.bookE..16M}
}

@article{Offner2023,
	adsurl = {https://ui.adsabs.harvard.edu/abs/2023ASPC..534..275O},
	archiveprefix = {arXiv},
	author = {{Offner}, S.~S.~R. and {Moe}, M. and {Kratter}, K.~M. and {Sadavoy}, S.~I. and {Jensen}, E.~L.~N. and {Tobin}, J.~J.},
	booktitle = {Protostars and Planets VII},
	editor = {{Inutsuka}, S. and {Aikawa}, Y. and {Muto}, T. and {Tomida}, K. and {Tamura}, M.},
	eprint = {2203.10066},
	month = jul,
	pages = {275},
	primaryclass = {astro-ph.SR},
	series = {Astronomical Society of the Pacific Conference Series},
	title = {{The Origin and Evolution of Multiple Star Systems}},
	volume = {534},
	year = 2023}

@article{Moe2017,
	adsurl = {https://ui.adsabs.harvard.edu/abs/2017ApJS..230...15M},
	archiveprefix = {arXiv},
	author = {{Moe}, Maxwell and {Di Stefano}, Rosanne},
	doi = {10.3847/1538-4365/aa6fb6},
	eid = {15},
	eprint = {1606.05347},
	journal = {Astrophys. J. Supp. S.},
	month = jun,
	number = {2},
	pages = {15},
	primaryclass = {astro-ph.SR},
	title = {{Mind Your Ps and Qs: The Interrelation between Period (P) and Mass-ratio (Q) Distributions of Binary Stars}},
	volume = {230},
	year = 2017}

@article{Kroupa,
	adsurl = {https://ui.adsabs.harvard.edu/abs/2001MNRAS.322..231K},
	archiveprefix = {arXiv},
	author = {{Kroupa}, Pavel},
	doi = {10.1046/j.1365-8711.2001.04022.x},
	eprint = {astro-ph/0009005},
	journal = {Mon. Not. R. Astron. Soc.},
	month = apr,
	number = {2},
	pages = {231-246},
	primaryclass = {astro-ph},
	title = {{On the variation of the initial mass function}},
	volume = {322},
	year = 2001}

@article{Liu2015,
	adsurl = {https://ui.adsabs.harvard.edu/abs/2015MNRAS.447..747L},
	archiveprefix = {arXiv},
	author = {{Liu}, Bin and {Mu{\~n}oz}, Diego J. and {Lai}, Dong},
	doi = {10.1093/mnras/stu2396},
	eprint = {1409.6717},
	journal = {Mon. Not. R. Astron. Soc.},
	month = feb,
	number = {1},
	pages = {747-764},
	primaryclass = {astro-ph.EP},
	title = {{Suppression of extreme orbital evolution in triple systems with short-range forces}},
	volume = {447},
	year = 2015}

@misc{wolfe2026binaryblackholespinpopulation,
      title={Binary-black hole spin population results may be driven by prior degeneracies}, 
      author={Noah E. Wolfe and Asad Hussain and Jack Heinzel and Salvatore Vitale},
      year={2026},
      eprint={2609.10753},
      archivePrefix={arXiv},
      primaryClass={gr-qc},
      url={https://arxiv.org/abs/2609.10753}, 
}

@ARTICLE{Boco2026,
       author = {{Boco}, Lumen and {Bosi}, Michele and {Sgalletta}, Cecilia and {Romagnolo}, Amedeo and {Mapelli}, Michela},
        title = "{Can current models predict the local black hole merger rate?}",
      journal = {arXiv e-prints},
         year = 2026,
        month = jun,
          eid = {arXiv:2606.02725},
        pages = {arXiv:2606.02725},
          doi = {10.48550/arXiv.2606.02725},
archivePrefix = {arXiv},
       eprint = {2606.02725},
 primaryClass = {astro-ph.HE},
       adsurl = {https://ui.adsabs.harvard.edu/abs/2026arXiv260602725B}
}

@ARTICLE{Broekgaarden2026b,
       author = {{Broekgaarden}, Floor S.},
        title = "{Lower Your Rates: On Claims of a Binary Black Hole Merger-Rate Crisis}",
      journal = {arXiv e-prints},
         year = 2026,
        month = jun,
          eid = {arXiv:2606.28515},
        pages = {arXiv:2606.28515},
          doi = {10.48550/arXiv.2606.28515},
archivePrefix = {arXiv},
       eprint = {2606.28515},
 primaryClass = {astro-ph.HE},
       adsurl = {https://ui.adsabs.harvard.edu/abs/2026arXiv260628515B}
}

@article{Liu:2024SEOBNRE,
    author        = {Liu, Xiaolin and Cao, Zhoujian and Zhu, Zong-Hong},
    title         = {Effective-one-body numerical-relativity waveform model
                     for eccentric spin-precessing binary black hole coalescence},
    journal       = {Class. Quant. Grav.},
    volume        = {41},
    number        = {19},
    pages         = {195019},
    year          = {2024},
    doi           = {10.1088/1361-6382/ad72ca},
    eprint        = {2310.04552},
    archivePrefix = {arXiv},
    primaryClass  = {gr-qc}
}

@article{Morras:2025nlp,
    author        = {Morras, Gonzalo and Pratten, Geraint and Schmidt, Patricia},
    title         = {Improved post-Newtonian waveform model for inspiralling
                     precessing-eccentric compact binaries},
    journal       = {Phys. Rev. D},
    volume        = {111},
    number        = {8},
    pages         = {084052},
    year          = {2025},
    doi           = {10.1103/PhysRevD.111.084052},
    eprint        = {2502.03929},
    archivePrefix = {arXiv},
    primaryClass  = {gr-qc}
}

@article{Albanesi:2025txj,
    author        = {Albanesi, Simone and Gamba, Rossella and
                     Bernuzzi, Sebastiano and Fontbut{\'e}, Joan and
                     Gonzalez, Alejandra and Nagar, Alessandro},
    title         = {Effective-one-body modeling for generic compact binaries
                     with arbitrary orbits},
    journal       = {Phys. Rev. D},
    volume        = {112},
    number        = {12},
    pages         = {L121503},
    year          = {2025},
    doi           = {10.1103/3snf-w1x7},
    eprint        = {2503.14580},
    archivePrefix = {arXiv},
    primaryClass  = {gr-qc}
}

@article{Morras:2026pyEFPEHM,
    author        = {Morras, Gonzalo and Pratten, Geraint and
                     Schmidt, Patricia and Buonanno, Alessandra},
    title         = {Post-Newtonian inspiral waveform model for eccentric
                     precessing binaries with higher-order modes and matter effects},
    journal       = {Phys. Rev. D},
    volume        = {114},
    number        = {4},
    pages         = {044032},
    year          = {2026},
    doi           = {10.1103/lxtg-6psv},
    eprint        = {2604.11903},
    archivePrefix = {arXiv},
    primaryClass  = {gr-qc}
}

@article{Gamboa:2026SEOBNRv6EPHM,
    author        = {Gamboa, Aldo and Pompili, Lorenzo and
                     Buonanno, Alessandra and Sebastiani, Luca and
                     Enficiaud, Raffi and Boyle, Michael and
                     Kidder, Lawrence E. and Pfeiffer, Harald P. and
                     Ramos-Buades, Antoni and Scheel, Mark A.},
    title         = {Enabling gravitational-wave astronomy with
                     spin-precessing black holes on generic orbits},
    year          = {2026},
    eprint        = {2609.01568},
    archivePrefix = {arXiv},
    primaryClass  = {gr-qc}
}

@ARTICLE{Vynatheya2022,
       author = {{Vynatheya}, Pavan and {Hamers}, Adrian S.},
        title = "{How Important Is Secular Evolution for Black Hole and Neutron Star Mergers in 2+2 and 3+1 Quadruple-star Systems?}",
      journal = {\apj},
         year = 2022,
        month = feb,
       volume = {926},
       number = {2},
          eid = {195},
        pages = {195},
          doi = {10.3847/1538-4357/ac4892},
archivePrefix = {arXiv},
       eprint = {2110.14680},
 primaryClass = {astro-ph.HE},
       adsurl = {https://ui.adsabs.harvard.edu/abs/2022ApJ...926..195V}
}

@ARTICLE{HamersDosopoulou2019,
       author = {{Hamers}, Adrian S. and {Dosopoulou}, Fani},
        title = "{An Analytic Model for Mass Transfer in Binaries with Arbitrary Eccentricity, with Applications to Triple-star Systems}",
      journal = {\apj},
         year = 2019,
        month = feb,
       volume = {872},
       number = {2},
          eid = {119},
        pages = {119},
          doi = {10.3847/1538-4357/ab001d},
archivePrefix = {arXiv},
       eprint = {1812.05624},
 primaryClass = {astro-ph.SR},
       adsurl = {https://ui.adsabs.harvard.edu/abs/2019ApJ...872..119H}
}

@article{Rodriguez2018cluster,
	adsurl = {https://ui.adsabs.harvard.edu/abs/2018PhRvD..98l3005R},
	archiveprefix = {arXiv},
	author = {{Rodriguez}, Carl L. and {Amaro-Seoane}, Pau and {Chatterjee}, Sourav and {Kremer}, Kyle and {Rasio}, Frederic A. and {Samsing}, Johan and {Ye}, Claire S. and {Zevin}, Michael},
	doi = {10.1103/PhysRevD.98.123005},
	eid = {123005},
	eprint = {1811.04926},
	journal = {Phys. Rev. D},
	month = dec,
	number = {12},
	pages = {123005},
	primaryclass = {astro-ph.HE},
	title = {{Post-Newtonian dynamics in dense star clusters: Formation, masses, and merger rates of highly-eccentric black hole binaries}},
	volume = {98},
	year = 2018}

@ARTICLE{Plunkett2026,
       author = {{Plunkett}, Cailin and {Callister}, Thomas and {Zevin}, Michael and {Vitale}, Salvatore},
        title = "{Signatures of a subpopulation of hierarchical mergers in the GWTC-4 gravitational-wave dataset}",
      journal = {arXiv e-prints},
         year = 2026,
        month = jan,
          eid = {arXiv:2601.07908},
        pages = {arXiv:2601.07908},
          doi = {10.48550/arXiv.2601.07908},
archivePrefix = {arXiv},
       eprint = {2601.07908},
 primaryClass = {gr-qc},
       adsurl = {https://ui.adsabs.harvard.edu/abs/2026arXiv260107908P}
}

@article{Rodriguez2019,
	adsurl = {https://ui.adsabs.harvard.edu/abs/2019PhRvD.100d3027R},
	archiveprefix = {arXiv},
	author = {{Rodriguez}, Carl L. and {Zevin}, Michael and {Amaro-Seoane}, Pau and {Chatterjee}, Sourav and {Kremer}, Kyle and {Rasio}, Frederic A. and {Ye}, Claire S.},
	doi = {10.1103/PhysRevD.100.043027},
	eid = {043027},
	eprint = {1906.10260},
	journal = {Phys. Rev. D},
	month = aug,
	number = {4},
	pages = {043027},
	primaryclass = {astro-ph.HE},
	title = {{Black holes: The next generation{\textemdash}repeated mergers in dense star clusters and their gravitational-wave properties}},
	volume = {100},
	year = 2019}

@ARTICLE{Flanagan2026a,
       author = {{Flanagan}, Elizabeth and {Antonini}, Fabio and {Callister}, Thomas and {Chattopadhyay}, Debatri and {Dosopoulou}, Fani and {Romero-Shaw}, Isobel and {Stegmann}, Jakob},
        title = "{Transitions in the Mass-ratio and Spin Properties of Binary Black Holes in GWTC-5}",
      journal = {arXiv e-prints},
         year = 2026,
        month = jun,
          eid = {arXiv:2606.14472},
        pages = {arXiv:2606.14472},
          doi = {10.48550/arXiv.2606.14472},
archivePrefix = {arXiv},
       eprint = {2606.14472},
 primaryClass = {astro-ph.HE},
       adsurl = {https://ui.adsabs.harvard.edu/abs/2026arXiv260614472F}
}

@ARTICLE{Belczynski2020,
       author = {{Belczynski}, K. and {Klencki}, J. and {Fields}, C.~E. and {Olejak}, A. and {Berti}, E. and {Meynet}, G. and {Fryer}, C.~L. and {Holz}, D.~E. and {O'Shaughnessy}, R. and {Brown}, D.~A. and {Bulik}, T. and {Leung}, S.~C. and {Nomoto}, K. and {Madau}, P. and {Hirschi}, R. and {Kaiser}, E. and {Jones}, S. and {Mondal}, S. and {Chruslinska}, M. and {Drozda}, P. and {Gerosa}, D. and {Doctor}, Z. and {Giersz}, M. and {Ekstrom}, S. and {Georgy}, C. and {Askar}, A. and {Baibhav}, V. and {Wysocki}, D. and {Natan}, T. and {Farr}, W.~M. and {Wiktorowicz}, G. and {Coleman Miller}, M. and {Farr}, B. and {Lasota}, J.-P.},
        title = "{Evolutionary roads leading to low effective spins, high black hole masses, and O1/O2 rates for LIGO/Virgo binary black holes}",
      journal = {\aap},
         year = 2020,
        month = apr,
       volume = {636},
          eid = {A104},
        pages = {A104},
          doi = {10.1051/0004-6361/201936528},
archivePrefix = {arXiv},
       eprint = {1706.07053},
 primaryClass = {astro-ph.HE},
       adsurl = {https://ui.adsabs.harvard.edu/abs/2020A&A...636A.104B}
}

@ARTICLE{Tong2026,
       author = {{Tong}, Hui and {Fishbach}, Maya and {Thrane}, Eric and {Mould}, Matthew and {Callister}, Thomas A. and {Farah}, Amanda M. and {Guttman}, Nir and {Banagiri}, Sharan and {Beltran-Martinez}, Daniel and {Farr}, Ben and {Galaudage}, Shanika and {Godfrey}, Jaxen and {Heinzel}, Jack and {Kalomenopoulos}, Marios and {Miller}, Simona J. and {Vijaykumar}, Aditya},
        title = "{Evidence of the pair-instability gap from black-hole masses}",
      journal = {\nat},
         year = 2026,
        month = apr,
       volume = {652},
       number = {8111},
        pages = {874-877},
          doi = {10.1038/s41586-026-10359-0},
archivePrefix = {arXiv},
       eprint = {2509.04151},
 primaryClass = {astro-ph.HE},
       adsurl = {https://ui.adsabs.harvard.edu/abs/2026Natur.652..874T}
}

@ARTICLE{Antonini2026,
       author = {{Antonini}, Fabio and {Romero-Shaw}, Isobel M. and {Callister}, Thomas and {Dosopoulou}, Fani and {Chattopadhyay}, Debatri and {Ginat}, Yonadav Barry and {Gieles}, Mark and {Mapelli}, Michela},
        title = "{Gravitational-wave constraints on the pair-instability mass gap and nuclear burning in massive stars}",
      journal = {Nature Astronomy},
         year = 2026,
        month = may,
          doi = {10.1038/s41550-026-02847-0},
archivePrefix = {arXiv},
       eprint = {2509.04637},
 primaryClass = {astro-ph.HE},
       adsurl = {https://ui.adsabs.harvard.edu/abs/2026NatAs.tmp..111A}
}

@ARTICLE{Flanagan2026b,
       author = {{Flanagan}, Elizabeth and {Stegmann}, Jakob and {Romero-Shaw}, Isobel and {Callister}, Thomas and {Olejak}, Aleksandra and {Antonini}, Fabio},
        title = "{Distinct spin properties and astrophysical origin of low mass binary black holes in gravitational wave data}",
      journal = {arXiv e-prints},
         year = 2026,
        month = jul,
          eid = {arXiv:2607.00565},
        pages = {arXiv:2607.00565},
archivePrefix = {arXiv},
       eprint = {2607.00565},
 primaryClass = {astro-ph.HE},
       adsurl = {https://ui.adsabs.harvard.edu/abs/2026arXiv260700565F}
}

@ARTICLE{GWTC5pop,
       author = {{The LIGO Scientific Collaboration} and {the Virgo Collaboration} and {the KAGRA Collaboration}},
        title = "{GWTC-5.0: Population Properties of Merging Compact Binaries}",
      journal = {arXiv e-prints},
         year = 2026,
        month = may,
          eid = {arXiv:2605.27226},
        pages = {arXiv:2605.27226},
          doi = {10.48550/arXiv.2605.27226},
archivePrefix = {arXiv},
       eprint = {2605.27226},
 primaryClass = {astro-ph.HE},
       adsurl = {https://ui.adsabs.harvard.edu/abs/2026arXiv260527226T}
}

@ARTICLE{Kapil2026,
       author = {{Kapil}, Veome and {Mandel}, Ilya and {Riley}, Jeff and {Grishin}, Evgeni and {Fuller}, Jim and {Berti}, Emanuele},
        title = "{Modern tidal interaction models for rapid binary population synthesis: II. Binary black hole formation, mergers, and spins}",
      journal = {arXiv e-prints},
         year = 2026,
        month = jun,
          eid = {arXiv:2606.23773},
        pages = {arXiv:2606.23773},
          doi = {10.48550/arXiv.2606.23773},
archivePrefix = {arXiv},
       eprint = {2606.23773},
 primaryClass = {astro-ph.HE},
       adsurl = {https://ui.adsabs.harvard.edu/abs/2026arXiv260623773K}
}

@article{Ray2026,
  author  = {Ray, Anarya and Mukherjee, Shirsha and Zevin, Michael and Kalogera, Vicky},
  title   = {On the Astrophysical Origin of Binary Black Hole Subpopulations: A Tale of Three Channels?},
  journal = {The Astrophysical Journal Letters},
  year    = {2026},
  volume  = {1005},
  number  = {2},
  pages   = {L55},
  doi     = {10.3847/2041-8213/ae80cb}
}

\end{document}